\documentclass[11pt,a4paper]{article}
\pdfoutput=1
\usepackage{subcaption}
\usepackage{jheppub}
\usepackage{extarrows}
\usepackage{caption}
\allowdisplaybreaks
\usepackage{graphicx}
\usepackage{subcaption}
\usepackage{slashed}
 \usepackage{setspace} 
 \usepackage{amsmath}
\usepackage{caption}
\usepackage{xcolor}  
\usepackage{color}
\allowdisplaybreaks
\def\bea{\begin{eqnarray}}
\def\eea{\end{eqnarray}}
\def\be{\begin{equation}}
\def\ee{\end{equation}}

\usepackage[utf8]{inputenc}

\DeclareUnicodeCharacter{2212}{-}
\usepackage{comment}
\usepackage[T1]{fontenc} 
\def\nue{{\nu_e}}
\def\anue{{\bar{\nu}_e}}
\def\numu{{\nu_{\mu}}}
\def\anumu{{\bar{\nu}_{\mu}}}

\newcommand{\eg}{{\it e.g.}}
\newcommand{\ie}{{\it i.e.}}

\newcommand{\beq}{\begin{equation}}
\newcommand{\eeq}{\end{equation}}
\newcommand{\beqa}{\begin{eqnarray}}
\newcommand{\eeqa}{\end{eqnarray}}

\usepackage{cancel}
\usepackage{soul}

\usepackage{amsmath,amssymb}
\usepackage{tikz}
\usepackage{tikz-feynman}
\tikzfeynmanset{compat=1.1.0}
\usepackage{slashed}
\usepackage{physics}
\usepackage{makecell}
\vspace*{-2cm}

\title{Implications for  new physics solutions to the LSND and  MiniBooNE  anomalies after MicroBooNE results}
  \author[a]{Waleed Abdallah,}
  \author[b]{Raj Gandhi}
  \author[c,d]{and Samiran Roy}
\affiliation[a]{Department of Mathematics, Faculty of Science, Cairo University, Giza 12613, Egypt}
\affiliation[b]{Harish-Chandra Research Institute, Chhatnag Road, Jhunsi, Allahabad 211019, India}
\affiliation[c]{Amity School of Engineering and Technology, Amity University Hyderabad, Hyderabad 500108, Telangana, India}
\affiliation[d]{School of Physics, University of Hyderabad, Hyderabad 500046, India}

    \emailAdd{awaleed@sci.cu.edu.eg} \emailAdd{raj@hri.res.in}
    \emailAdd{samiranroy.hri@gmail.com}
  \abstract{
  Over the past couple of decades, the Liquid Scintillator Neutrino Detector (LSND) and MiniBooNE (MB)  neutrino experiments have provided statistically significant evidence of persistent and as yet unexplained electron-like event excesses. The MicroBooNE experiment, with sensitive particle identification capabilities designed to shed light on these anomalous results, has conducted a number of analyses to test whether the excess is due to photons, electrons or $e^+e^-$ pair productions, all of which give almost identical signals in LSND and MB. Recent MicroBooNE searches, in addition to checking Standard Model (SM) processes, have focused on proposed new physics scenarios involving new mediators and dark portals beyond the SM.
  We study the consequences and effects of these analyses for the many theoretical proposals put forth to generate the three types of signals above in order to understand the observed excesses. We discuss which classes of solutions are allowed, disallowed or require further scrutiny involving theoretical and experimental efforts. We give an example of a new solution that survives the restrictions imposed by MicroBooNE results and other existing bounds. We point out that a light ($m\leq 30$~MeV) dark neutral spin zero  mediator  with dominantly pseudoscalar couplings, decaying to an $e^+e^-$ pair, can account for the observations. We indicate that if such a model is realised, it may point to the possibility that neutrinos are Dirac particles, and that there are dark Heavy Neutral Leptons (HNLs) of mass ${\cal O}(100)$~MeV. We discuss issues related to CP violation and ultra-violet completion of these models. Finally we stress  that a search for the excess in incoherent scattering channels in MicroBooNE and SBND could help clarify the physics behind these persistent anomalies.
}
\begin{document}
\maketitle
\flushbottom 
%%%%%%%%%%%%%%%%%%%%%%% Section 1 %%%%%%%%%%%%%%%%%%%%%%%%%
\section{Introduction}
%%%%%%%%%%%%%%%%%%%%%%%%%%%%%%%%%%%%%%%%%%%%%%%%%%%%%%%%%%%
The search for new but weakly coupled physics Beyond the Standard Model (BSM) at low energy scales has received added impetus in recent years,  due to the lack of any non-standard physics signals at the high energies of the  Large Hadron Collider (LHC). Possible fertile environments for discoveries in the low energy regime include neutrino and Dark Matter (DM) detectors, due to their being optimized to pick up signals of weak (\ie, feeble) interactions at small energies.

Over the last couple of decades, this line of inquiry has in particular been fed by results from the Liquid Scintillator Neutrino Detector (LSND)~\cite{LSND:2001aii} and MiniBooNE (MB)~\cite{MiniBooNE:2007uho,MiniBooNE:2008yuf,MiniBooNE:2013uba}, both of which are short-baseline liquid scintillation detectors with incident $\numu$ and/or $\anumu$ beams. Average beam energies for both experiments are below 1~GeV, and both have reported excesses in electron-like events\footnote{The term ``electron-like events'', as used in this paper comprises three categories of final states. Both MB and LSND, being mineral-oil based detectors, were only sensitive to the total amount of visible light deposited in any given interaction, and consequently could not distinguish between electrons, photons or (somewhat collimated) electron-positron pairs.}. The significances of the LSND and MB excesses are $3.8\sigma$ and  $4.8\sigma$ respectively and their combined significance stands at $6.1\sigma$. Over the years,  careful checks and studies of standard expected backgrounds have been conducted (see~\eg~\cite{LSND:1996jxj, Katori:2020tvv, Dasgupta:2021ies, Brdar:2021ysi, Alvarez-Ruso:2021dna, MicroBooNE:2021zai}) in order to eliminate SM physics explanations.

Earlier work on understanding the LSND and MB excesses in terms of physics BSM was based largely on the existence of one or more light ($m\sim$~eV) sterile neutrinos. Such states, with appropriate masses and mixings with the SM neutrinos, naturally lead to additional $\numu\rightarrow \nue$ or $\anumu \rightarrow \anue$ oscillations which can then be used to understand the signals in these two detectors. This hypothesis was also applied towards understanding other low-energy anomalies~\cite{GALLEX:1994rym, GALLEX:1997lja, Kaether:2010ag, Abdurashitov:1996dp, SAGE:1998fvr, Abdurashitov:2005tb, SAGE:2009eeu} which involved electron-neutrino  disappearance at short baselines. The existence of such light inactive neutrinos is, however, currently in strong tension with cosmology~\cite{Hamann:2011ge, Archidiacono:2013xxa, Hagstotz:2020ukm}, which limits {\it{i}}) the number of relativistic degrees of freedom in thermal equilibrium prior to neutrino decoupling at $T\sim 1$~MeV as well as {\it{ii}})  the amount of hot DM in the Universe. Additionally, it exhibits significant tension with $\numu$ disappearance results~\cite{MINOS:2020iqj,IceCube:2020phf,IceCube:2020tka}. Inconsistencies and tensions are also manifest in $\nu_e$ disappearance data~\cite{Giunti:2021kab}. These diverse disagreements  adversely affect global analyses and fits of neutrino data~\cite{Dentler:2018sju,Diaz:2019fwt,Boser:2019rta,Dasgupta:2021ies,Acero:2022wqg} which attempt to incorporate this hypothesis.

This squeeze on the light sterile neutrino hypothesis has motivated a significant amount of work on other new physics explanations of the MB and LSND anomalies. In this context, an important constraint from MB, running in the beam-dump mode, has helped shrink the realm of scenarios to be considered~\cite{MiniBooNEDM:2018cxm}. During normal running, protons were incident on a beryllium target, producing charged pions which decayed to neutrinos and anti-neutrinos in the  region immediately downstream of the target. In the beam-dump mode, the target was by-passed and the protons were directed to the steel absorber at the end of the decay region. This led to a strong suppression of the neutrino flux\footnote{Charged mesons produced in a thin target like beryllium escape the target and produce decay-in-flight neutrinos. In the beam-dump mode, with a thick absorber, these mesons are either absorbed or decay at rest within a few radiation lengths, reducing the neutrino flux significantly. Neutral mesons, however,  decay electromagnetically in flight due to their short lifetimes in both the normal and beam-dump modes.}, but made the detector sensitive (due to the suppression of the background caused by neutrino scattering in the detector) to DM scattering and its decay. If the observed excess were due to a process occurring in the beam dump, \eg, DM production, instead of neutrino-related processes, the predicted excess would scale with the amount of Protons On Target (POT) collected.  An example of such a process would be the production of a dark mediator through neutral meson decay or proton bremsstrahlung, which would then decay into two DM particles, $\chi_1$ and $\chi_2$. One of these  could reach the detector,  producing an excess of electron-like events via processes like  $\chi_{1,2} + e \rightarrow \chi_{1,2} + e $, $\chi_2\rightarrow \chi_1 + e^+e^-$ or $\chi_2\rightarrow \chi_1 +\gamma$~\cite{Batell:2009di,deNiverville:2011it,Kahn:2014sra,deNiverville:2016rqh,Izaguirre:2017bqb,Jordan:2018gcd}.

No excess or scaling with POT was observed, \ie, the anomalous events were not present when the neutrino production was curtailed. This led to the conclusion that the excess was directly linked to the production, propagation or detection of neutrinos, narrowing the possibilities for new physics scenarios.

   Ideas to understand the excesses rely on the ambiguity mentioned earlier in the observation of final states in LSND and MB. Such proposals have typically involved, in various combinations, the introduction of new mediators and/or Heavy Neutral Leptons (HNLs) and transition magnetic moments, which lead to signals in the detectors that may explain the excess (for a review see~\cite{Abdullahi:2023ejc}). One class of proposals~\cite{Moss:2017pur,Moulai:2019gpi, Akhmedov:2011zza, Bramante:2011uu, Karagiorgi:2012kw, Asaadi:2017bhx, Smirnov:2021zgn, Alves:2022vgn, Palomares-Ruiz:2005zbh, Bai:2015ztj, deGouvea:2019qre, Dentler:2019dhz, Hostert:2020oui,Chang:2021myh} combines sterile oscillations and/or decay with new interactions. A second category focuses on non-oscillatory interactions using these elements of new physics, which produce an excess of electron-like events inside the detectors~\cite{Gninenko:2009ks,Gninenko:2010pr,Gninenko:2012rw,Masip:2012ke,Radionov:2013mca,Magill:2018jla,Bertuzzo:2018itn,Ballett:2018ynz,Ballett:2019pyw,Datta:2020auq,Dutta:2020scq,Abdallah:2020biq,Abdullahi:2020nyr,Abdallah:2020vgg,Schwetz:2020xra,Vergani:2021tgc,Hammad:2021mpl,Dutta:2021cip,Alvarez-Ruso:2021dna,Abdallah:2022grs,Kamp:2022bpt,Bansal:2022zpi,Ghosh:2023dgk,Abdallah:2024uby,Dutta:2025fgz}.

Over the past few years, the MicroBooNE experiment~\cite{MicroBooNE:2016pwy}, an 85 metric ton active volume Liquid Argon Time Projection Chamber (LArTPC) situated in the Booster Neutrino Beam (BNB) at Fermilab, has added important components to the interpretation and understanding of the anomalies. Specifically, MicroBooNE has been designed to study the MiniBooNE anomaly as its primary goal. Both detectors were fed by the same neutrino beam and had similar baselines. Its recent result~\cite{MicroBooNE:2025nll} uses data obtained in its LArTPC from two accelerator neutrino beams to exclude the single light sterile neutrino interpretation of the LSND and MiniBooNE anomalies at the 95\% Confidence Level (CL)\footnote{A distinction must be made between scenarios relying solely on active-sterile oscillations and those which combine such oscillations in vacuum or matter along with aspects of new interactions~\cite{Moss:2017pur,Moulai:2019gpi, Akhmedov:2011zza, Bramante:2011uu, Karagiorgi:2012kw, Asaadi:2017bhx, Smirnov:2021zgn, Alves:2022vgn, Palomares-Ruiz:2005zbh, Bai:2015ztj, deGouvea:2019qre, Dentler:2019dhz, Hostert:2020oui,Chang:2021myh}. The latter category requires a more focussed case-specific examination to determine viability.}. In addition, it rules out a notable portion of the sterile neutrino parameter space that could explain the gallium anomaly~\cite{Kaether:2010ag,SAGE:2009eeu,Barinov:2022wfh}.

Earlier MicroBooNE results include ({\it{i}}\,) previous searches for an anomalous excess of electron-like events~\cite{MicroBooNE:2021tya, MicroBooNE:2021pvo, MicroBooNE:2021nxr,MicroBooNE:2024tym}, ({\it{ii}}\,) a direct search using the BNB  for a light oscillating sterile neutrino~\cite{MicroBooNE:2022sdp}, and 
{ ({\it{iii}}\,)} a search for Neutral Current (NC) production of the $\Delta(1232)$ baryon resonance followed by $\Delta$ radiative decay~\cite{MicroBooNE:2021zai}.  These searches have recently been followed by a search for NC single coherent photons~\cite{MicroBooNE:2025rsd}, an inclusive single photon search~\cite{MicroBooNE:2025ntu} and a search for dark sector production of $e^+e^-$ pairs~\cite{MicroBooNE:2025khi}.

Given these results, it is both germane and timely to examine their implications for the new physics solutions mentioned above which involve neutrinos (excluding those that rely on active-sterile oscillations alone), and to specifically attempt to understand which types of solutions are disfavored and which ones, being still allowed, warrant further experimental and theoretical scrutiny. The next section (Section~\ref{Sec:muB})  is devoted to this effort. Section~\ref{Sec:Interaction} presents a new example of an allowed solution involving a pseudoscalar mediator and the consequent fits it provides to the LSND and MB data. Section~\ref{Sec:constraints} discusses constraints, as well as certain theoretical and phenomenological topics relevant to those solutions that are still allowed, pointing out where further investigation may be warranted. The final section summarizes our conclusions.
%%%%%%%%%%%%%%%%%%%%%%% Section 2 %%%%%%%%%%%%%%%%%%%%%%%%%
\section{Recent MicroBooNE results and their implications for new physics solutions of the LSND and MB anomalies}
\label{Sec:muB}
%%%%%%%%%%%%%%%%%%%%%%%%%%%%%%%%%%%%%%%%%%%%%%%%%%%%%%%%%%%
We begin with some general remarks.  The MB experiment was built and designed to try to verify and possibly understand the LSND anomaly. Its results, while compatible with the anomalous excess of electron-like events seen by LSND, have added aspects to the puzzle that 
are new. If the origins of the anomalies under discussion here lie in new physics, it is important that such physics simultaneously address and encompass the anomalous observations in both detectors. Many proposed new physics solutions focus on explaining MB alone. It is possible that such attempts may not provide a complete picture of the physics necessary to understand the observations because attempting to explain both sets of results simultaneously imposes certain requirements on the new physics~\cite{Abdallah:2022grs} which do not arise when one attempts to understand either MB or LSND alone\footnote{This statement, of course, assumes that both anomalies are genuine in the sense that they cannot be explained by SM backgrounds or systematics.}.  

Secondly, prior to discussing specific searches, we note that MicroBooNE can, and does, distinguish between single electrons and photons (both of which give electromagnetic showers in the detector), using two methods: $a)$ measuring the $dE/dx$, which, for a photon is expected to be approximately twice as much as that for an electron, and $b)$  looking for a gap ($25$ cm on the average) between the vertex and the inception of the shower in events where there is a vertex-identifying accompanying proton. 

\subsection{Electron searches}
As mentioned above, MicroBooNE conducted multiple searches~\cite{MicroBooNE:2025nll,MicroBooNE:2021tya, MicroBooNE:2021nxr,MicroBooNE:2021wad,MicroBooNE:2024tym} for any electrons that could account for the observed excesses. The searches simultaneously tested the hypothesis that an increase in the $\nu_e$ rate led to the anomalous events, and ruled out such an explanation at $> 99\%$ CL. The results provide constraints which would apply to the light sterile neutrino hypothesis, as well as any new physics scenarios~\cite{Moss:2017pur,Moulai:2019gpi, Akhmedov:2011zza, Bramante:2011uu, Karagiorgi:2012kw, Asaadi:2017bhx, Smirnov:2021zgn, Alves:2022vgn, Berryman:2018ogk, Palomares-Ruiz:2005zbh, Bai:2015ztj, deGouvea:2019qre, Dentler:2019dhz, Hostert:2020oui} which increase the $\nu_e$ or the electron rate in the detectors above what is nominally expected. 

\subsection{Photon searches}
An important set of searches carried out by MicroBooNE looked at neutrino-induced NC resonant $\Delta (1232)$ baryon production followed by $\Delta$ radiative decay. Below $\sim 1$~GeV, this is expected to be the dominant source of single photon events predicted in both neutrino-argon (relevant for MicroBooNE) and neutrino-carbon (relevant for MB and LSND) scattering~\cite{Wang:2013wva}. It selected single photon events with one or zero protons and without charged leptons in the final state ($1\gamma 1p$ and $1\gamma 0p$, respectively).

The search constrained its background using an in-situ high-purity measurement of NC $\pi^0$ events, made possible via dedicated $2\gamma 1p$ and $2\gamma 0p$ selections. In total, 16 and 153 events were observed for the $1\gamma 1p$ and $1\gamma 0p $ cases, respectively, in comparison with a constrained background prediction of $20.5\pm3.65$ (sys.) and $145.1\pm13.8$ (sys.) events. These results set a bound on an anomalous enhancement of the normalization of NC $\Delta$ radiative decay to be less than 2.3 times the predicted SM rate at the 90$\% $ CL. Consequently, it disfavored a photon interpretation of the MB low-energy excess as a factor of 3.18 times the expected SM  NC $\Delta$ radiative decay rate at the 94.8$\%$~CL. 

Indirectly, the measurement thus constrains new physics models which would enhance the $\Delta$ production and decay rate to single photon channels in order to explain the observed excesses. An example of such a model would be one that postulates an additional $Z'$-boson which achieves this via its mixing with the SM $Z$-boson. We note, however, that the $1\gamma 1p$ measurement relied on the observation of a proton in conjunction with the single photon final state, and also used the fact that the invariant mass of the photon-proton pair was $\sim 1232$~MeV. This handle was unavailable for the $1\gamma 0p$ event sample. The analysis of zero-proton events is consequently more challenging. The nominal SM expectation was used, where the ${\rm BR}(\Delta^+\rightarrow p + \gamma)\equiv {\rm BR}(\Delta^0\rightarrow n+ \gamma)$. Also, the analysis indicated that while the MB excess could not be explained by mostly $1\gamma 1p$ events, the $1\gamma 0p$ events were compatible both with the expected SM rate as well as a possible excess. Overall, the results of this search cannot exclude  an interpretation of the
MB excess  as being primarily due to an enhancement of $1\gamma 0p$ NC events which could arise via appropriate new physics which does not concurrently enhance the $1\gamma 1p$ events.

A subsequent search focused on NC coherent photons~\cite{MicroBooNE:2025rsd}. In the SM, this is an extremely rare process, with a rate that is about $10\%$ of the resonant $\Delta (1232)$ production.
It involves the interaction of a neutrino with the whole nucleus at once while
leaving it in its nuclear ground state and producing a
single photon with no hadronic activity. The purpose of the search was not discovery as such, since the rate in MicroBooNE was expected to be very low, but it hoped to see any anomalous excess of coherent photons that could conceivably explain the observations. Coherent photons produced via the rare SM interaction were expected to be low in energy, and fairly forward with respect to the neutrino beam direction. While MicroBooNE required a proton to have a minimum of 35~MeV in kinetic energy for its other searches in order to reconstruct the proton, it used a proton stub veto which eliminated even very low-energy protons for the coherent gamma search. Uncertainties were constrained using NC $\pi^0$ events, and overall, no significant coherent gamma excess was observed.

MicroBooNE also conducted an inclusive single gamma search~\cite{MicroBooNE:2025ntu}. Both the NC resonant $\Delta (1232)$ events as well as the coherent single gamma events discussed above were subsets of the total expected sample in this search. In addition, however, the search encompassed any single photon production which could arise as a common background to other searches, $\eg$  a $\pi^0$ decay where one of the two photons was $a$) below the detection threshold of $20$~MeV, or $b$) had exited the detector prior to detection or $c)$ was one of a pair produced just outside the detector but failed to enter it. Two co-linear photons with an opening angle less than $20^{\circ}$ would also be identified as a single photon in this search.

Importantly, the search was designed to pick up single photon events originating in new physics processes which could, $\eg$  arise from the decay of HNLs or dark neutrinos to a photon. HNL decay to a low-energy co-linear $e^+e^-$ pair could also be misidentified and detected as a single photon. (This last channel has implications for the dark $e^+e^-$ search, which we discuss below.)

Overall, the results of the inclusive single photon search showed that the full sample, encompassing all single photon events of both SM and BSM origins,  was consistent with expectations within $1.6 \sigma$. A smaller sub-sample, that with zero protons, however, showed a $2 \sigma$ excess in events below $E_{\rm shower}\simeq 600$~MeV. We note that since there was an imposed proton energy threshold of $E\geq 35$ MeV, part of the excess in the zero proton sample could very well arise due to $1\gamma1p$ events which have a low energy proton. This, along with the non-observation of any excesses in exclusive channels ($\eg$ radiative $\Delta$ decay, or the coherent gamma sample) forms the gist of the photon search results of MicroBooNE. New physics proposals which are based on excess gamma production~\cite{Gninenko:2009ks, Gninenko:2010pr, Gninenko:2012rw, Masip:2012ke, Radionov:2013mca, Magill:2018jla, Datta:2020auq, Schwetz:2020xra, Vergani:2021tgc, Alvarez-Ruso:2021dna, Kamp:2022bpt, Bansal:2022zpi, Dutta:2025fgz} must conform to and be tested against  these results.  

\subsection{Search for dark sector decay to $e^+e^-$}

One of the possibilities that arises from the lack of any single electron excess, in conjunction with a mild ($\sim 2\sigma$) excess in the $1\gamma 0p$ channel below $600$~MeV may be the production of low-energy $e^+e^-$ pairs due to the decay of a dark particle. Such a signal would be indistinguishable for the most part in MicroBooNE from a single SM  photon due to the lack of an identifiable vertex. This motivated a wider search for $e^+e^-$ pairs~\cite{MicroBooNE:2025khi}, which yields important restrictions on new physics proposals as we discuss below.

The search in~\cite{MicroBooNE:2025khi} focused on very forward-going, zero-proton, coherently produced $e^+e^-$ events over a wide range of energies and opening angles.  A large number of such events were expected in new physics scenarios~\cite{Bertuzzo:2018itn,Ballett:2018ynz,Ballett:2019pyw,Abdullahi:2020nyr,Abdallah:2020biq,Hammad:2021mpl} proposing a new $Z'$ mediator, which could either be light ($\eg\; 30$~MeV~\cite{Bertuzzo:2018itn}) or relatively heavy ($\eg\; 1.25$~GeV~\cite{Ballett:2018ynz}). It encompassed both i) single dark neutrino topologies  where the incoming $\nu_{\mu}$ up-scattered to a heavier dark neutrino, which then decayed to a $Z'$ and to a light active neutrino with the subsequent production of an $e^+e^-$ pair and ii) double dark neutrino topologies where the first dark neutrino decayed to a lighter dark neutrino and a $Z'$. In these interactions, the $Z'$ could decay to an $e^+e^-$ pair  both on-shell or off mass-shell, depending on its mass. (For more details on signal selection, background rejection,  pair topologies covered by the search,  energy and angle dependence of efficiencies etc, see~\cite{MicroBooNE:2025khi}.)

Overall, the search saw no evidence of a coherent $e^+e^-$ signal which 
was predicted by models with a new vector mediator and single or double
dark neutrinos. It thus  constrained the scenarios proposed in~\cite{Ballett:2018ynz,Ballett:2019pyw,  Abdullahi:2020nyr,Abdallah:2020biq,Hammad:2021mpl}. It consequently provided the first direct limits on these dark sector models and, at the $95\%$ CL, excluded the majority of the parameter space that they proposed as a solution to the MB anomaly.  We note that  LSND observations also indicate that the excess did not reside in coherent events, because their anomalous events were always accompanied by a neutron. This could be an indirect indication that the same new physics could be responsible for the excesses observed in  both~detectors.

\subsection{Overall conclusions from the above searches}

The results from the three search categories above have narrowed the plethora of options that were available as possible explanations of the MB and LSND excesses prior to MicroBooNE results. 

One important possibility which remains and invites further investigation are models with dark neutrinos/HNLs and scalar and/or pseudoscalar mediators~\cite{Abdallah:2020vgg, Abdallah:2024uby} which allow for the excess to be produced largely via incoherent scattering, leading to the production of $e^+e^-$ pairs. Scalar and pseudoscalar mediators provide certain advantages over vectors with reference to the problem at hand, viz. simultaneously understanding the excesses in LSND and MB. The relevant cross-section, as discussed in~\cite{Abdallah:2022grs} allows enough events to be produced at LSND
(which had lower energies) once model parameters are fixed to fit MB results. This is difficult to achieve with vector mediators due to the rapid fall of the cross-section as one goes down in energy in the range relevant to MB and LSND.
Secondly, as energy increases above this range, the cross-section for vectors rises in a way that conflicts~\cite{Arguelles:2018mtc} with limits
from CHARM II~\cite{CHARM-II:1994dzw} and MINERvA~\cite{MINERvA:2019hhc}. Scalar and pseudoscalar mediators evade these constraints since the cross-section falls rapidly with increasing energy beyond the range relevant to MB and LSND.

The lack of a significant number of coherent single photon events as well as coherently produced $e^+e^-$ pairs were essential to ruling out new physics scenarios which rely on vector mediators. This problem can be ameliorated by scalar mediators. The real scalar scenario of~\cite{Abdallah:2020vgg} for instance, yields only $\sim 15\%$ coherently produced pairs versus about $\sim 85\%$ incoherently produced ones. In this model, which has both a light singlet scalar and a heavier one as part of a  SM additional doublet, the contribution from the light scalar is predominantly coherent. However, its overall effect remains small due to its small  mixing angle with the heavier scalar. In contrast, the contribution from the heavy scalar is largely incoherent. A scenario with a pseudoscalar mediator, as in~\cite{Abdallah:2024uby} or like the one described in the next section would give only incoherent events because the mediator couples to the net spin of the target, which is zero for the nuclei under consideration (Carbon in LSND and MB and Argon in MicroBooNE).

Next, we mention a couple of points pertinent to the interpretation of whether a particular event is a photon or a pair. MicroBooNE was unable to distinguish between a nearly co-linear pair of $e^+e^-$ and a single photon. Thus, the $2 \sigma$ excess seen in the zero-proton sample below 600~MeV (discussed above) could have either of these possibilities as its origin. Moreover, as mentioned earlier, protons with energies $\leq 35$ MeV fall below the detection threshold. Thus a $1\gamma0p$ event could very well be a collimated $e^+e^-$ pair event with a low energy proton. Secondly, with reference to the search for pairs, we note from~\cite{MicroBooNE:2025khi} that detection efficiencies for pairs with low invariant mass (as in models proposed in~\cite{Abdallah:2020vgg, Abdallah:2024uby}) are low and any excess residing in them could very well have been missed. For a low mass pair, higher energies would lead to significant collimation, and thus would likely be misidentified as a single photon, while low energy configurations would fall below detection thresholds. 

 As also emphasized in a previous sub-section, given the results of the $1\gamma0p$ search, it is certainly possible that the excess does reside in a scenario that produces a single gamma~\cite{Gninenko:2009ks, Gninenko:2010pr, Gninenko:2012rw, Masip:2012ke, Radionov:2013mca, Magill:2018jla, Datta:2020auq, Schwetz:2020xra, Vergani:2021tgc, Alvarez-Ruso:2021dna, Kamp:2022bpt, Bansal:2022zpi, Dutta:2025fgz}, however, it should not simultaneously significantly enhance coherent gamma production or $1\gamma1p$ events beyond nominal expectations. The proposals above thus need to be checked against these criteria established by MicroBooNE.

 Finally, (as also briefly mentioned above) we comment on the non-observation of any electron-like event excess in coherent events in LSND, \ie~there was no excess in events where a single electron-like event unaccompanied by any hadronic activity was seen.  The observed excess resided solely in (incoherent) events with an accompanying neutron. Assuming that the same new physics is responsible for excesses in both LSND and MB, we see that this apparent suppression of coherent electron-like events in LSND is consistent with the non-observation of such events in  MicroBooNE in its efforts to explain the MB excess.
%%%%%%%%%%%%%%%%%%%%%%% Section 3 %%%%%%%%%%%%%%%%%%%%%%%%%
 \section{The Trilinear Pseudoscalar Interaction}
\label{Sec:Interaction}
%%%%%%%%%%%%%%%%%%%%%%%%%%%%%%%%%%%%%%%%%%%%%%%%%%%%%%%%%%
In this section, we present a new example of a model with a pseudoscalar mediator which can produce the excess in both LSND and MB, and which provides very good fits to both the energy and angular distributions observed in those detectors.

We consider an extension of the SM including a new pseudoscalar field $a'$ 
that couples to a muon neutrino $\nu_\mu$, a single HNL $N$, 
and to the SM quarks $q$. The gauge invariant Lagrangian (given below)  also has a trilinear self-interaction term for the pseudoscalar.

\subsection{The model}

 The scalar sector of the SM is expanded by incorporating a second Higgs doublet, and, as mentioned above,  we also add a singlet pseudoscalar $\phi_{h'}=i \, A_3^0/\sqrt{2}$. Additionally, one HNL $N$ participates in the interaction which generates electron-like signals in MB and LSND.  We can write the scalar potential $V$ as
\begin{eqnarray}
    V=V_{\rm 2HDM}+V_{h'},
\end{eqnarray}
where $V_{\rm 2HDM}$ and $V_{h'}$ are given in the Higgs basis $(\phi_h,\phi_H,\phi_{h'})$ as follows:
\begin{eqnarray}
V_{\rm 2HDM}&=& \mu_1 |\phi_{h}|^2 +\mu_2 |\phi_{H}|^2 -\mu_3\big{\{}\phi_{h}^\dagger\phi_{H}+h.c.\big{\}}+ \frac{\lambda_1}{2} |\phi_{h}|^4+ \frac{\lambda_2}{2} |\phi_{H}|^4 +\lambda_3 |\phi_{H}|^2|\phi_{h}|^2\nonumber\\
&&\!\!\!\!\!\!\!\!+\lambda_4 (\phi_{h}^\dagger\phi_{H})(\phi_{H}^\dagger\phi_{h})
+\frac{\lambda_5}{2} \big{\{}(\phi_{h}^\dagger\phi_{H})^2 +h.c.\big{\}}+ (\lambda_6 |\phi_{h}|^2 + \lambda_7  |\phi_{H}|^2)\, (\phi_{h}^\dagger\phi_{H}+ \phi_{H}^\dagger\phi_{h})\nonumber,\\[0.1cm]  
V_{h'}&=&\mu' |\phi_{h'}|^2+m'\phi_{h'}^3 + \lambda'_2 |\phi_{h'}|^4+\lambda'_3 |\phi_{h}|^2 |\phi_{h'}|^2+\lambda'_4|\phi_{H}|^2 |\phi_{h'}|^2  \nonumber\\
&& +\big{\{}\lambda'_5 |\phi_{h'}|^2(\phi_{h}^\dagger\phi_{H}) +(m_1|\phi_{h}|^2+m_2|\phi_{H}|^2+m_3\phi_{h}^\dagger\phi_{H}- m_s \phi_{h'})\phi_{h'}+h.c. \big{\}}\nonumber.
\label{eq:lagrangian}
\end{eqnarray}
Here,
\begin{eqnarray} 
\phi_{h}&=&\left( \begin{array}{c}
H^+_1 \\
\frac{v+H^0_1+i A^0_1}{\sqrt{2}} \\
\end{array} \right),~~~
\phi_{H}=\left( \begin{array}{c}
H^+_2 \\
\frac{H^0_2+i A_2^0}{\sqrt{2}} \\
\end{array} \right),~~~
\phi_{h'}=i\, A_3^0/\sqrt{2}\,.
\end{eqnarray}
We consider the vacuum expectation values (VEV) $\langle\phi_h\rangle=v(\equiv v_{\rm SM})\simeq  246$~GeV and $\langle \phi_H\rangle\!=\!0\!=\!\langle \phi_{h'}\rangle$. Here, $H^+_1, A^0_1$ are the Goldstone modes, which give the gauge bosons mass after the electroweak symmetry is spontaneously broken.

The mass matrix of the neutral CP-even Higgses in the basis $\left(H_1^0, H_2^0\right)$ is given by
\begin{equation} \label{CP-Even-MM}
{\cal M}^2_{\cal H} = \left( 
\begin{array}{ccc}
\lambda_1 v^2 &\lambda_6 v^2  \\ 
\lambda_6 v^2 & \mu_H 
 \end{array} 
\right), 
 \end{equation} 
where $\mu_H = \mu_2 +(\lambda_3 + \lambda_4 + \lambda_5)v^2/2$.  We have minimized the scalar potential $V$ using the following conditions:
\begin{equation}
\mu_1 = -\frac{1}{2}\lambda_1 v^2,~~\mu_{3} = \frac{1}{2}\lambda_6 v^2.
\label{eq:conds}
\end{equation} 
The matrix in eq.~\eqref{CP-Even-MM},   $m^2_{\cal H}$, is diagonalized by $Z^{\cal H}$ as follows:
\begin{equation} 
Z^{\cal H} {\cal M}^2_{\cal H} (Z^{\cal H})^T = ({\cal M}^{2}_{\cal H})^{\rm diag}\,, 
~~{\rm with}~~
H^0_i = \sum_{j}Z_{{j i}}^{\cal H}h_{{j}}\,,
\end{equation} 
\begin{equation}
{Z^{\cal H}}=\left(
\begin{array}{ccc}
  c_\alpha & -s_\alpha \\
  s_\alpha & c_\alpha
\end{array}
\right),~~\tan{2 \alpha}=\frac{ 2\lambda_6 v^2 }{\lambda_1 v^2 -\mu_H},
\end{equation}
where $s_\alpha\equiv \sin \alpha$, $c_\alpha \equiv \cos \alpha$. In the alignment limit ($\lambda_6\sim0$), the SM-like Higgs is $H^0_1\approx h$ with $M^2_h\simeq \lambda_1 v^2$ and $H^0_2\approx H$ having mass $M_H^2\simeq\mu_H$.
The mass matrix of the neutral CP-odd Higgses in the basis $\left(A_2^0, A_3^0\right)$, satisfying the  conditions in eq.~\eqref{eq:conds}, is given~by
 \begin{equation} \label{CP-Odd-MM}
{\cal M}^2_{\cal A} = \left( 
\begin{array}{ccc}
 \mu_A & -m_{3} v/\sqrt{2}  \\ 
 -m_{3} v/\sqrt{2}  &\mu_{a'}
 \end{array} 
\right), 
 \end{equation} 
where, $\mu_A = \mu_H- \lambda_5 v^2$ and $\mu_{a'}=\mu'+ 2\,m_s +\lambda'_3 v^2/2$. The  masses of the  CP-odd physical Higgs states $(A, a')$ are given by
\begin{equation}
M^{2}_{A,a'}\simeq\frac{1}{2}\left[\mu_A+\mu_{a'}\!\pm\! \sqrt{(\mu_A-\mu_{a'})^2 + 2 m_{3}^2 v^2 }\right]\!.
\label{eq:masscpeven2}
\end{equation}
 The  corresponding mixing matrix and angle $\xi$ is given by
\begin{equation}
{Z^{\cal A}}=\left(
\begin{array}{ccc}
c_\xi & -s_\xi \\
s_\xi & c_\xi
\end{array}
\right),~~\tan{2 \xi}=\frac{\sqrt{2}\, m_{3} v}{\mu_A-\mu_{a'}}.
\end{equation}
 The charged Higgs mass is given by 
\begin{eqnarray}
M^2_{H^\pm}&=&\mu_2+\lambda_3 v^2/2\,.
\label{eq:mass2}
\end{eqnarray}
In the Higgs basis the relevant Lagrangian ${\cal L}$ can be written as follows
\begin{eqnarray}
{\cal L}&=&\sqrt{2}\,\Big[(X^u_{ij} \tilde\phi_h+\bar{X}^u_{ij} \tilde\phi_H) \bar{Q}_L^i  u^j_R
+(X^d_{ij}  \phi_h+\bar{X}^d_{ij} \phi_H) \bar{Q}_L^i  d^j_R+(X^e_{ij}  \phi_h+\bar{X}^e_{ij} \phi_H) \bar{L}_L^i  e^j_R\nonumber\\
&+&(X^\nu_{i}  \tilde\phi_h +\bar{X}^\nu_{i} \tilde\phi_H) \bar{L}_L^i  N_R+\frac{1}{\sqrt{2}} M_{N} \bar{N}_L N_R+\lambda_N\bar{N}_L\phi_{h'}N_R+h.c.\Big],
\label{eq:YkN}
\end{eqnarray}
here, $X^f_{ij}~(f=u,d,e)$ stands for SM Yukawa couplings depending on SM charged fermion mass, whereas $\bar{X}^f_{ij}$ (with $i,j=1,2,3$) are independent Yukawa matrices.  The fermion masses receive contributions only from $X^f_{ij}~(f=u,d,e)$, since  we work in the Higgs basis, \ie, only $\phi_h$ acquires a non-zero VEV while $\langle\phi_H\rangle$ = $\langle\phi_{h'}\rangle=0$, leading to $X^f=\frac{M_f}{v}$, where $M_f$ are the SM charged fermion mass matrices. We note the absence of the right-handed counterparts $\nu_R$ of the SM neutrino fields $\nu_L$ in the Lagrangian above. This has been done by  invoking a $Z_2$
 symmetry under which $\nu_{R}\rightarrow -\nu_{R}$, keeping all other fields even. In effect, we have opted to make the SM neutrinos, as well as $N$, to be Dirac particles, by imposing lepton number conservation but simultaneously eliminating un-naturally small Yukawa couplings. We justify this choice in the next section.
$\bar{X}^f_{ij}$ are free parameters and non-diagonal matrices. We  work in a basis in which charged fermion mass matrices are real and diagonal, where $U_k M_f V_k^\dagger=m_f^{\rm diag}=\frac{y^h_f}{\sqrt{2}} \, v$ are the requisite bi-unitary transformations. The additional independent doublet sector Yukawa couplings $\bar{X}^f_{ij}$ are henceforth considered in a diagonal basis (as $y_f$) for simplicity. Finally, after a rotation of the scalar fields, one finds the following coupling strengths of the scalars $h$, $H$,  $A$, and $a'$ with fermions:
\begin{eqnarray}
y^h_f= \frac{\sqrt{2} \, m_f}{v},  ~ y^H_f= y_f , ~ y^A_f = \, y_f  \, c_\xi,~ y^{a'}_f =  \, y_f  \, s_\xi,  ~ \lambda^A_{N} = \,\lambda_N \, s_\xi, ~{\rm and }~ \lambda^{a'}_{N }= \, \lambda_N  \, c_\xi\,. \, \, \, \, \, \, \,
\label{eq:fermcouplingstrength1}
\end{eqnarray}

\subsection{The interaction in MB and LSND}

The relevant renormalizable interaction Lagrangian is
\begin{equation}
-\mathcal{L} \supset
\kappa\,a'^3+y^{a'}_q\,a'\,\bar q\,\gamma_5\, q  +(y^{a'}_{\nu_\mu N}\,a'\,\bar{N}\,\gamma_5\,P_L\,\nu_\mu
 + h.c.) \,,
\end{equation}
where \(P_L=\tfrac{1}{2}(1-\gamma_5)\) and $\kappa$, in our model, is given by:
\begin{equation}
\kappa=\frac{3}{\sqrt{2}} \Big{[}m'c_\xi^3+\frac{v}{\sqrt{2}}\,(\lambda'_5-\lambda'^*_5)c_\xi^2s_\xi+\left(m_2^*-m_2\right) s_{\xi }^2c_{\xi }+\frac{v}{\sqrt{2}}\,(\lambda^*_7-\lambda_7)s^3_\xi\Big{]}\, \,.
\end{equation}
In case of $\lambda'_5, m_2$ and $\lambda_7$ are real, $\kappa$ becomes:
\begin{equation}\label{Eq:kappa-approx}
\kappa=\frac{3}{\sqrt{2}} m'c_\xi^3\, \,.
\end{equation}
\begin{figure}[t!]
\centering
\begin{tikzpicture}
  \begin{feynman}
    \vertex (n1) at (-2,1.5) {$\nu_\mu(k_1)$};
    \vertex (n2) at (2,1.5) {$N(k_2)$};
    \vertex (vtop) at (0,1.5);
    \vertex (q1) at (-2,-1.5) {$n(p_1)$};
    \vertex (q2) at (2,-1.5) {$n(p_2)$};
    \vertex (vbot) at (0,-1.5);
    \vertex (vc) at (0,0);
    \vertex (aout) at (2,0) {$a'(k)$};

    \diagram* {
      (n1) -- [fermion] (vtop) -- [fermion] (n2),
      (q1) -- [fermion] (vbot) -- [fermion] (q2),
    };

    \draw[dashed] (vtop) -- node[right] {$a'(q_\nu)$} (vc);
    \draw[dashed] (vbot) -- node[right] {$a'(q_n)$} (vc);
    \draw[dashed] (vc) -- (aout);

% Vertices
\vertex (a) at (4,0);
\vertex (v) at (6,0);
\vertex (b) at (8,1) {$e^-$};
\vertex (c) at (8,-1) {$e^+$};

% Diagram
\diagram* {
(a) -- [scalar, dashed, edge label=$a'$] (v),
(v) -- [fermion] (b),
(v) -- [anti fermion] (c),
};
    
  \end{feynman}
\end{tikzpicture}

\caption{Feynman diagram of the scattering process via the trilinear pseudoscalar coupling, which leads to the  MB and LSND excesses
 in our model.}
\label{Fig:FD-MB}
\end{figure}
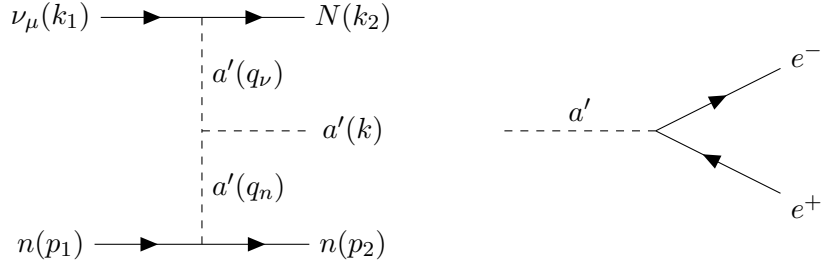
The scattering process of interest (see figure~\ref{Fig:FD-MB}) is
\begin{equation}
\nu_\mu + n \;\longrightarrow\; N + n + a' (a'\to e^+e^-) \,,
\end{equation}
where $n$ denotes either a neutron or a proton. It is assumed that the $N$ either escapes the detector prior to decaying, or decays to invisible particles. The matrix element for this process and derivation of the differential cross-section $(d\sigma/dE_{a'}^{\rm lab})$, including the 3-body phase space and kinematic integration limits, are presented in Appendix~\ref{app}. For calculations of the total differential cross-section and number of events in MB and LSND, see~\cite{Abdallah:2024uby}.

The goodness of fit is quantified using the chi-squared test, which is given by
\begin{equation}
\chi^2
=
\sum_i
\frac{\left(N_i^{\rm obs}-N_i^{\rm pred}\right)^2}
{\sigma_{\rm{stat},i}^2+\sigma_{\rm{syst},i}^2},
\end{equation}
where $N_i^{\rm obs}$ and $N_i^{\rm pred}$ denote the observed and predicted numbers of events in the $i$th bin, respectively. Assuming Poisson statistics, the statistical uncertainty is approximated by the observed number of events ($\sigma_{\rm{stat},i}^2\simeq N_i^{\rm obs}$). The MB collaboration has provided only the systematic uncertainty in the energy distribution~\cite{MiniBooNE:2020pnu}, therefore for both the energy and angular distributions we use an uncorrelated fractional systematic uncertainty of $13\%$ to be consistent with the fractional systematic uncertainty reported in Ref.~\cite{MiniBooNE:2020pnu}, \ie, $\sigma_{\rm{syst},i}^2=(0.13 \,N_i^{\rm obs})^2$ for the MB distributions. For the LSND distributions we assume a $15\%$  fractional systematic uncertainty.
%%%%%%%%%%%%%%%%%%%%%%%%%%%
\begin{figure}[t]
\begin{center}
\includegraphics[width=10cm,height=7.5cm]{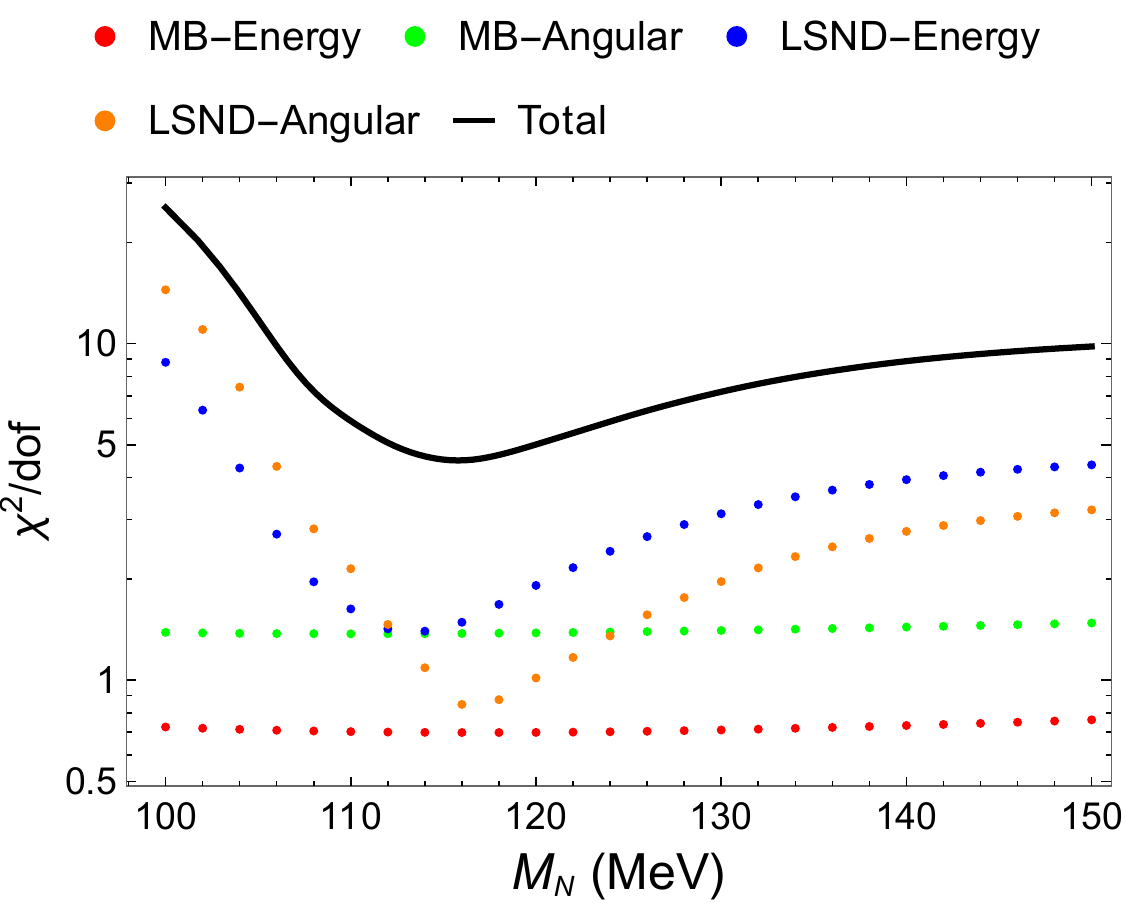}
\caption{Reduced chi-squared, $\chi^2/\mathrm{dof}$, as a function of the heavy neutrino mass $M_N$. The colored points show the individual contributions from the MB energy (red), MB angular (green), LSND energy (blue), and LSND angular (orange) distributions, while the black curve represents the total reduced chi-squared obtained by summing all contributions.}
\label{Fig:MB-LSND-chi2dof}
\end{center}
\end{figure}
%%%%%%%%%%%%%%%%%%%%%%%%%%%
 The total chi-squared $\chi^2_{\rm tot}$ is obtained by summing the contributions from the MB and LSND energy and angular distributions.

To keep the quality of the fit independent of the number of data points, we use the chi-squared per degree
of freedom (dof),
$\chi^2/\mathrm{dof}$ (also called the reduced chi-squared). Here the number of $\mathrm{dof}$ is the difference between 
the total number of fitted data points and the number of free parameters in the fit. A value of $\chi^2/\mathrm{dof}\simeq1$ indicates an excellent agreement between the model and the data, while significantly larger values correspond to increasingly poor fits.

\begin{figure}[!t]
\begin{center}
\includegraphics[width=7.5cm,height=6cm]{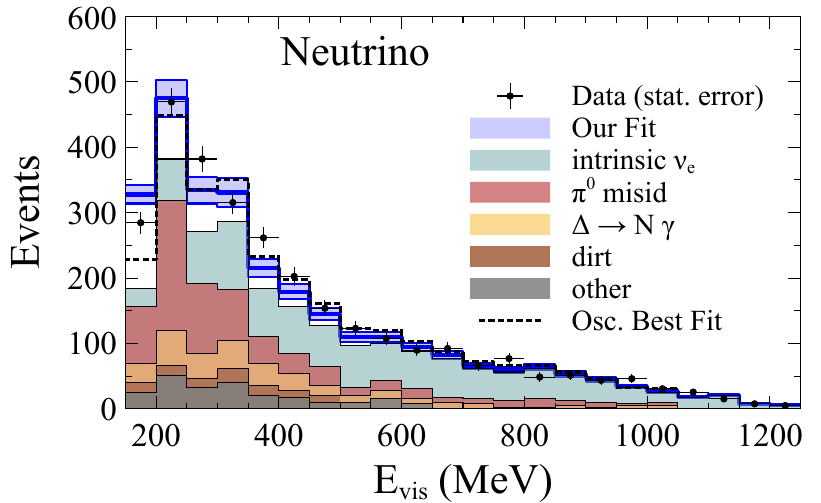}~\includegraphics[width=7.5cm,height=6cm]{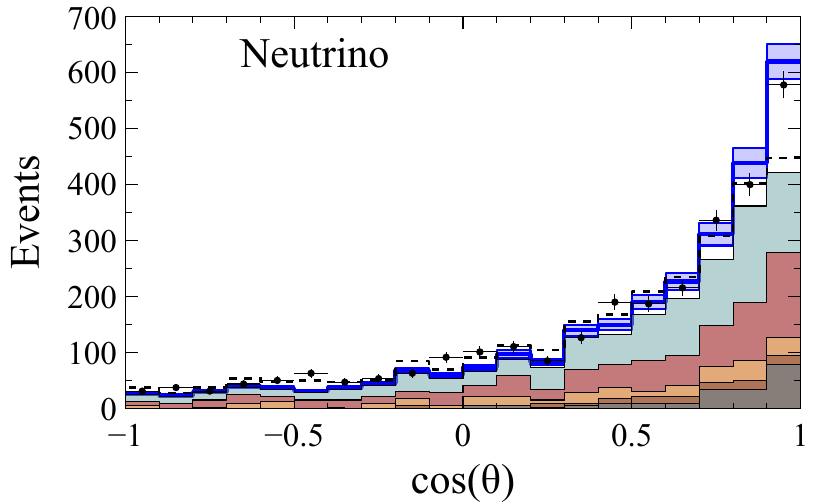}\\
\includegraphics[width=7.5cm,height=6cm]{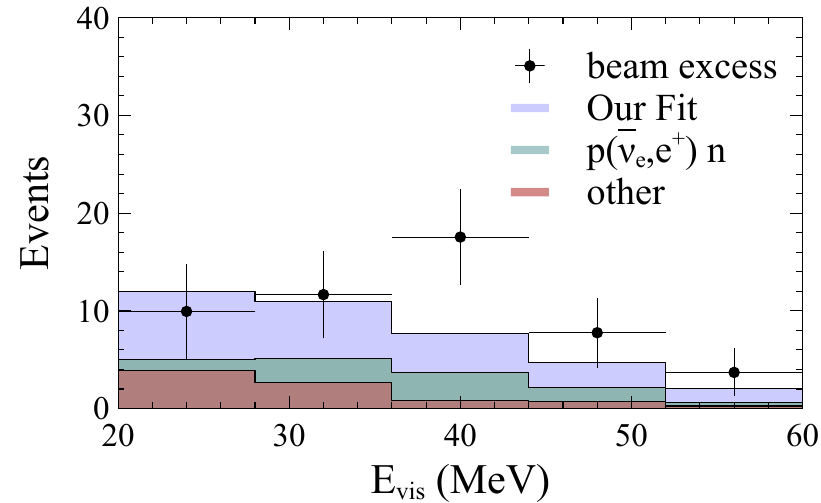}~\includegraphics[width=7.5cm,height=6cm]{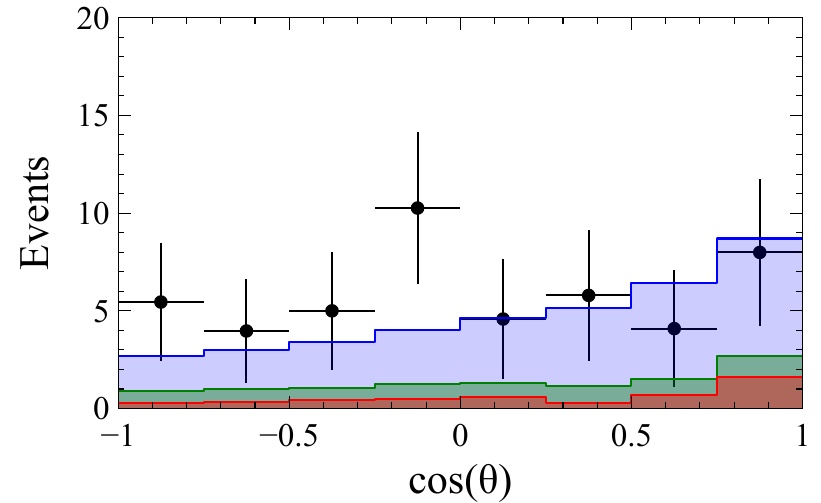}
\caption{Top panels: These show, for neutrino runs, the MB electron-like events (backgrounds and signal). They are  plotted against the visible energy E$_{\rm vis}$ (left panel) and  the emitted angle of the Cerenkov light (right panel). The blue solid line denotes the central prediction of excess events in our model.
Bottom panels: These show the energy distribution (left panel) of the LSND data the angular distribution (right panel) of the light originating from the electron-like final state. In both panels, the shaded blue region represents our fit with an  uncertainty band, and other shaded regions denote the backgrounds. We have used the values $M_N=114$~MeV, $M_{a'}$=17~MeV from Table~\ref{tab1}.}
\label{Fig:MB-LSND-dist}
\end{center}
\end{figure}

%%%%%%%%%%%%%%%%%%%%%%%%%%%
\begin{table}[hb]
\begin{center}
 \begin{tabular}{|c|c|c|c|c|c|}
  \hline
  \makecell{$M_{N}$\\[-0.05cm] (MeV)}& \makecell{$M_{a'}$\\[-0.05cm] (MeV)} &$\kappa$ &$y_{u(d)}^{a'}\times10^6$ & $y_{e}^{a'}\times 10^{5}$ & $y_{\nu_\mu N}^{a'}\times 10^{2}$  \\
  \hline
   114 & $17$ & $0.7068$& $4.75\,(4)$& $2.3$ & $5$ \\
  \hline
 \end{tabular}
\caption{Benchmark parameter values used for event generation in MB and LSND.}
\label{tab1}
\end{center}
\end{table}
%%%%%%%%%%%%%%%%%%%%%%%%%%%
Figure~\ref{Fig:MB-LSND-chi2dof} shows the individual and total values of $\chi^2/\mathrm{dof}$ as functions of the heavy neutrino mass $M_N$, with the remaining parameters fixed to the values given in Table~\ref{tab1}. The global fit exhibits a clear minimum around $M_N\simeq114~\mathrm{MeV}$, indicating the preferred mass region. While the MB energy and angular distributions provide nearly constant contributions over the entire mass range ($\chi^2_{\rm MB}/{\rm dof}\simeq0.7,1.4$, respectively), the LSND energy and angular distributions have their minimum reduced chi-squared values ($\chi^2_{\rm LSND}/{\rm dof}\simeq1.4,1.1$, respectively) at $M_N=114~\mathrm{MeV}$ and they become increasingly disfavored for $M_N\gtrsim120~\mathrm{MeV}$, leading to a steady increase in the total reduced chi-squared. Consequently, the overall fit quality is primarily driven by the LSND data at larger values of $M_N$. Table~\ref{tab1} lists the optimal values obtained by the analysis described here. They have been used to generate the event rates shown in figure~\ref{Fig:MB-LSND-dist}.

%%%%%%%%%%%%%%%%%%%%%%% Section 4 %%%%%%%%%%%%%%%%%%%%%%%%%
\section{Discussion on constraints and future directions}
\label{Sec:constraints}
%%%%%%%%%%%%%%%%%%%%%%%%%%%%%%%%%%%%%%%%%%%%%%%%%%%
In the example provided above (and in similar models)\footnote{Considerations discussed in this section may be applied, with small modifications, to other light scalar mediated models $\eg$ those in ~\cite{Abdallah:2020vgg,Abdallah:2024uby}.} several important constraints need to be conformed to. Beam-dump/fixed-target/missing-energy searches (NA64~\cite{NA64:2018lsq}, E137~\cite{Bjorken:1988as}, E141~\cite{Riordan:1987aw}, ORSAY~\cite{Davier:1989wz}) set strong limits on a pseudoscalar $a'$ that couples to electrons: roughly $\leq$ ~few ×$10^{-4}$ to $10^{-3}$ for $M_{a'} \simeq 17$~MeV (NA64~\cite{NA64:2018lsq}
quotes $\leq 6.8\times10^{-4}$ at $16.7$~MeV for a generic coupling to $e$).
Rare meson decays ($\pi^0 \rightarrow a' + X, K \rightarrow \pi + a', \pi^+\rightarrow e^+\nu e^+e^-$, etc.)~\cite{SINDRUM:1986klz} typically give constraints on couplings to light quarks and nucleons. Constraints also arise from measurements of the anomalous magnetic moments of the electron and muon. For a discussion of these and other constraints on such scalar and pseudoscalar models the reader is referred to~\cite{Abdallah:2020vgg, Abdallah:2024uby} and references therein.

 In what follows,  we discuss certain other important bounds and theoretical considerations which come into play  in light scalar mediator models attempting to explain MB and LSND.\\

{\textit{\textbf {CP violating  effects:}}}\\

In models with additional scalars, CP violating~(CPV)  considerations need to be carefully taken into account. We first take a general approach towards  CPV emanating in the scalar sector in models similar to ours (SM + 2HDM + light singlet pseudoscalar), \eg~ the model considered in~\cite{Abdallah:2024uby}. We present some sample calculations to illustrate the points we make. Where appropriate, we indicate the simplifications we have made to mitigate large CPV contributions  in the particular model considered above .  
 
  Once you have explicit or spontaneous CPV, the model enters tightly constrained territory. For a light pseudoscalar, the dominant CP-violating constraints are not collider bounds but precision, low-energy observables. A good example of the latter is the electron Electric Dipole Moment~(eEDM). Strong constraints arise from the ACME experiments~\cite{ACME:2018yjb,ACME:2016sci}, which restrict the eEDM $|{d_e}| < 1.1\times 10^{-29}~e\,{\rm cm}$. Generally speaking, in a model with mixing between the singlet and doublet pseudoscalars, as well as explicit or spontaneous CPV in the scalar sector, contributions to EDMs can be potentially large~\cite{Pospelov:2005pr,
Engel:2013lsa,Inoue:2014nva,Brod:2013cka,Gunion:2005ja}
arising from terms like
 \[
\mathcal{L} \supset y^\varphi_f \, \bar{f} \left( \cos\delta + i \gamma^5 \sin\delta \right) f \, \varphi +h.c.\,.
\]
\begin{figure}[t]
    \centering
        \begin{subfigure}[b]{0.32\textwidth}
        \centering
       \includegraphics[width=0.95\textwidth]{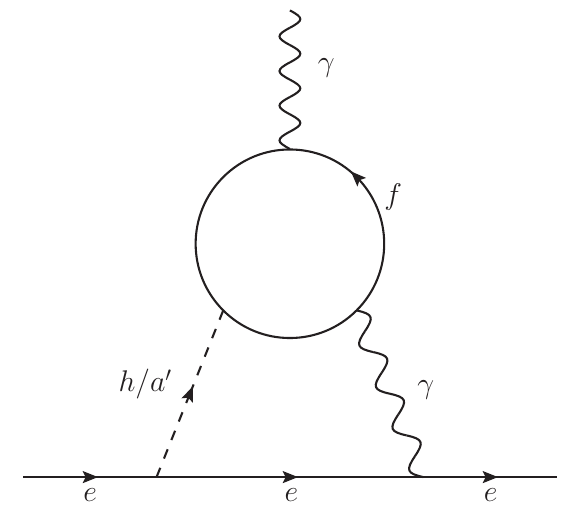}
        \caption{}
        \label{Fig:FD-BZ-2L}
    \end{subfigure}
    \hfill
    \begin{subfigure}[b]{0.32\textwidth}
        \centering
         \includegraphics[width=0.95\textwidth]{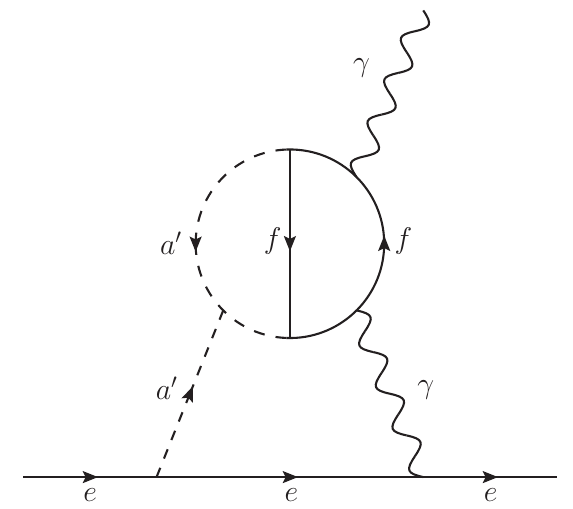}
        \caption{}
        \label{Fig:FD-BZ-3L}
    \end{subfigure}
    \hfill
    \begin{subfigure}[b]{0.32\textwidth}
        \centering
         \includegraphics[width=0.95\textwidth]{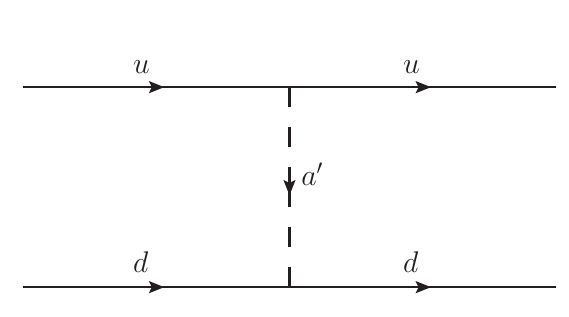}
        \caption{}
        \label{Fig:FD-4fn}
    \end{subfigure}
    
  \caption{Sample Feynman diagrams of dipole moment  contributions. (a)~Two-loop Barr-Zee diagram with the SM fermions $f$ running in the loop. The same diagram with the $W$-boson and charged Higgses instead of $f$ can contribute to the eEDM. (b)~Three-loop Barr-Zee diagram through the trilinear coupling of $a'$. (c)~Four-quark diagram contributing to the neutron dipole moment.}
    \label{Fig:FD-eEDM}
\end{figure}
This requires both singlet and doublet mixing with some angle $\theta$ along with CP-even states $(h,H)$ and CP-odd states $(a',A)$ mixing with angle $\theta_{\rm CP}$, because $\sin\delta \sim \theta\theta_{\rm CP}$. Such mixings lead to mass eigenstates which are not eigenstates of CP, and to contributions to two-loop Barr-Zee diagrams~\cite{Barr:1990vd} with  the SM fermions, the $W$-boson and charged Higgses (see figure~\ref{Fig:FD-BZ-2L}), leading to large EDM contributions.

\begin{figure}[!t]
\begin{center}
\includegraphics[width=7.3cm,height=6.2cm]{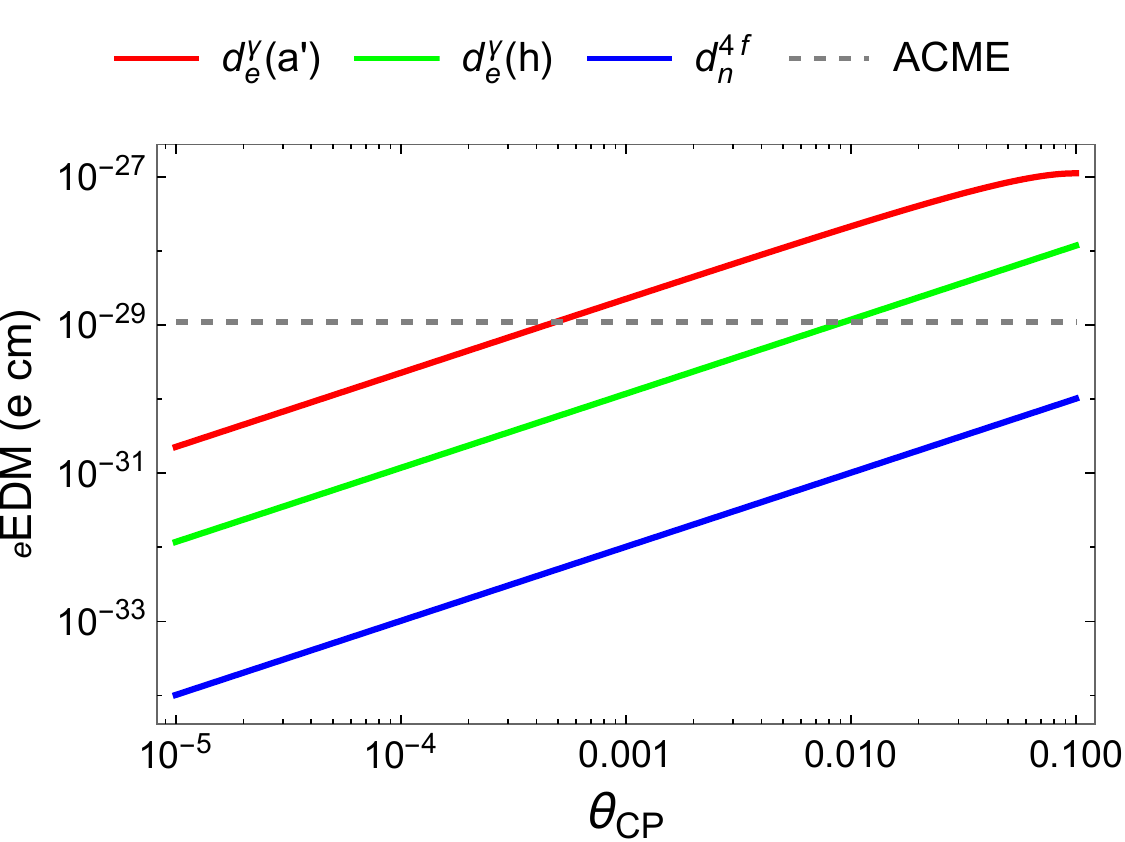}~~\includegraphics[width=7.5cm,height=6cm]{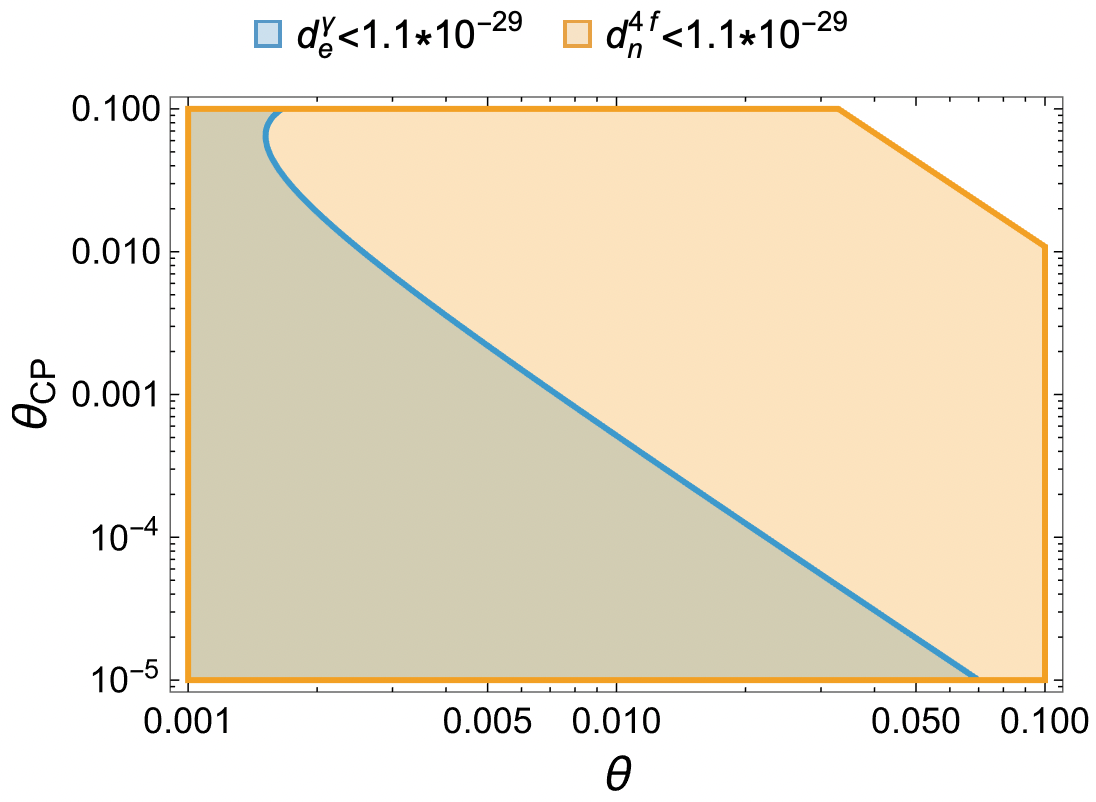}
\caption{Left: The eEDM contributions from the $a'$, $h$ and neutron diagrams versus the CP phase angle $\theta_{\rm CP}$. The horizontal gray dashed line refers to the ACME upper bound ($1.1\times 10^{-29}~e\,{\rm cm}$). Right: Regions allowed by the ACME upper bound in the $\theta-\theta_{\rm CP}$ plane.}
\label{Fig:eEDM}
\end{center}
\end{figure}
The relevant scalar states that contribute dominantly to such diagrams
 are the SM Higgs ($h$) and the light pseudoscalar ($a'$), and their contributions are denoted by $d_e^\gamma(h)$ and $d_e^\gamma(a')$, respectively. However, the $a'$ contribution to the two-loop Barr-Zee diagrams is larger than the SM Higgs contribution  by two orders of magnitude, as shown in left panel of figure~\ref{Fig:eEDM}. The reason is that the $a'$ coupling with the SM light fermions in our model, in particular with the electron pair, is enhanced by a factor $v^2/m_e^2$. For our benchmark point at $\theta=10^{-2}$, the eEDM approximately equals $2.3\times 10^{-26}\,\theta_{\rm CP}~e\,{\rm cm}$ and consequently $\theta_{\rm CP}$ has to be smaller than $5\times 10^{-4}$ to satisfy the ACME upper bound ($1.1\times 10^{-29}~e\,{\rm cm}$), as mentioned above. We have also calculated the four-quark contribution for the neutron ($d_n^{4f}$) to the EDMs of atoms and molecules~\cite{Jung:2013hka}, see figure~\ref{Fig:FD-4fn}. We find that it is smaller than the sum of $a'$ and $h$ contributions to the two-loop Barr-Zee diagrams ($d_e^\gamma=d_e^\gamma(h)+d_e^\gamma(a')$) by three orders of magnitude, as shown in left panel of figure~\ref{Fig:eEDM}. In the right panel of figure~\ref{Fig:eEDM}, we show the region allowed by the ACME upper bound. For the neutron EDM $d_n^{4f}$ this is shown  in orange and   for the two-loop Barr-Zee $d_e^\gamma$ it has been shown  in blue. We see  that the latter provides the most restrictive constraint on the $\theta-\theta_{\rm CP}$ plane, resulting in a significantly reduced allowed region compared to that obtained from the neutron EDM bound.

For  the specific model considered here,
(eq.~(\ref{eq:lagrangian})) these constraints require
$\theta_{\rm CP}$ to be extremely tiny or zero. (Note that a non-zero $\theta$ is necessary for the excess events to be produced.) Given current ACME limits, $\theta_{\rm CP} \leq 5\times 10^{-4}$ is sufficient to comply with the eEDM constraints. For the most part, in this work we make choices that automatically preserve CP in the terms (with one necessary exception).
 Mass eigenstates are kept either CP-even or CP-odd, by assuming $\theta_{\rm CP}=0$ as discussed above. Terms linear in $\phi_{h'}$, which has odd parity, can lead to potentially large Barr-Zee contributions, hence we choose $m_1, m_2$ and $m_3$ to be zero. Generically, choosing all couplings to be real is a safe choice. The term $m'\phi_{h'}^3$, is explicitly CP violating but is necessary since the trilinear interaction forms the basis of generating the excess events in our model. However, such a term gives suppressed contributions to the EDMs, $\eg$, contributions arise at three loops (see figure~\ref{Fig:FD-BZ-3L}) instead of two loops, and is likely to comfortably evade existing bounds.

 We also note that the ATLAS Collaboration has recently performed its first search for CP-violating contributions to the Higgs-Vector boson-Vector boson (HVV) vertex using the full LHC Run-3 dataset of $164$~fb$^{-1}$ at  $\sqrt s = 13.6$ TeV, exploiting the $h(125)\rightarrow \gamma\gamma$ decay in vector-boson-fusion production~\cite{ATLAS:2026ovz}. No significant deviation from the SM was observed. In the context of the model presented here (\ie~a singlet pseudoscalar + a 2HDM), we note that such a vertex would receive CPV contributions from the $\theta_{\rm CP}$ admixture of the Higgs with a pseudoscalar. The lack of a signal is in conformity with a very small 
 value of $\theta_{\rm CP}$ that we assume here based on the EDM bounds.\\

{\textit{\textbf {One-loop corrections to SM neutrino masses:}}}\\

A second consideration that requires further examination in scalar mediator models with HNLs, such as those considered in this work and in~\cite{Abdallah:2020vgg, Abdallah:2024uby}, are one-loop corrections to the masses of active neutrinos. Primarily, these would arise if the active neutrinos and HNLs (coupling to the singlet scalar) are Majorana particles. The corrections are larger for light singlet scalar and HNL  masses, and if the Yukawa couplings between the singlet scalar and the HNLs are large. If one assumes that the SM neutrinos and the HNLs are Dirac particles, loop corrections are usually benign. Specifically, a loop diagram requires a chirality flip which originates from lepton-number violation via a Majorana insertion, but in the  Dirac case no such insertion exists and 
  $\delta m = 0$ exactly at the lowest order. This is a motivating reason for our choice of neutrinos as Dirac particles in our model.

 While we postpone a fuller study to future work~\cite{Abdallah:2026xxx} we make some qualitative remarks regarding the possible mass generation mechanisms in such a scenario. The unexplained  smallness of the Dirac neutrino Yukawa couplings (in the SM)  compared to other fermions is a major motivation for Majorana neutrinos and  the seesaw mechanism. In principle, as has been explored extensively in the literature (for a review and references see~\cite{Chulia:2021jgv}), nature could have chosen other ways to address this problem which is inherent to Dirac neutrinos. For instance,  one could eliminate the SM Higgs  Yukawa couplings for neutrinos  by invoking a $Z_2$
 symmetry under which $\nu_{R}\rightarrow -\nu_{R}$,  and all other fields are even\footnote{This has been implemented in eq.~(\ref{eq:YkN}).}. SM neutrinos could get their small Dirac masses radiatively.  Specific manifestations  and ideas related to obtaining small Dirac neutrino masses  have been explored in~\cite{Roncadelli:1983ty,Dick:1999je,Gu:2006dc,Ma:2006km,Farzan:2012sa, CentellesChulia:2019gic,Ma:2016mwh,Bonilla:2018ynb,Calle:2018ovc,Cai:2017jrq}.

To summarize, in models such as ours,   the smallness of the SM neutrino Dirac masses can be  naturally explained by  loop suppression,  the imposition of an appropriate $Z_2$  symmetry and by appropriately including additional particles, \eg~as in the  Dirac scotogenic mechanism~\cite{Farzan:2012sa, Ma:2016mwh}. Unlike the Majorana case, lepton number conservation forbids the dangerous one-loop correction to SM neutrino masses from the $a',N$ loop~\cite{Calle:2018ovc,Cai:2017jrq}.\\

{\textit{\textbf {Considerations related to ultra-violet (UV) completion:}}}\\

The model presented here is an effective field theory valid below some UV cutoff scale $\Lambda_{\rm UV}$.
 Like other scenarios with  2HDMs, it inherits the standard hierarchy problem of the SM scalar sector.  Additionally, the independent Yukawa couplings $\bar{X}_{ij}$ of the second doublet require a UV mechanism — such as a flavor symmetry — to suppress potentially dangerous FCNCs.
 A second concern  is the 17~MeV mass of the singlet pseudoscalar $a'$, which is many orders of magnitude below the electroweak scale, and unprotected by any symmetry against radiative corrections.

 The UV concerns outlined  above are, however,  substantially ameliorated if $a'$
is considered to be  part of an independent dark sector which has  its own gauge invariance and particle content. In this picture, the smallness of $M_{a'} \sim 17$~MeV
 relative to the electroweak scale could possibly  reflect a  dynamical scale of the dark sector, $\Lambda_{\rm dark} \sim \mathcal{O}(20~{\rm MeV})$, similar to the manner in which $\Lambda_{\rm QCD}$
 sets the pion mass in the SM. If $a'$
 is a pseudo-Nambu-Goldstone boson of a spontaneously broken dark sector symmetry, its mass is technically natural in the  't Hooft sense, since it is protected by the approximate symmetry and generated dynamically by dark sector confinement or symmetry breaking. Radiative corrections to $M_{a'}$
from SM fields are suppressed by the dark sector gauge invariance and by the smallness of the portal couplings, keeping $M_{a'}$ stable against large corrections.  Since the full particle spectrum of the dark sector is unknown, we regard the model presented here as the low-energy effective description of a richer dark sector, whose complete UV structure is left for future investigation.

It must be noted, however, that unlike the mass of $a'$, the flavor structure of the independent second-doublet Yukawa matrices $\bar{X}_{ij}$ cannot be 
  ameliorated by the dark sector, as these couplings belong to the visible sector. In the absence of a protecting symmetry,  off-diagonal  elements of $\bar{X}_{ij}$
can generate tree-level FCNCs. In the present work we adopt a diagonal basis for $\bar{X}_{ij}$
as a simplifying assumption. A complete UV explanation of their flavor structure — via $\eg$ an explicit flavor symmetry — is an important open question that we leave for future work. We note, however,  that the decoupling of the heavy doublet states (with masses in the several hundred GeV range, as required by our constraints) provides a practical suppression of FCNC effects regardless of the detailed structure of $\bar{X}_{ij}$.\\

{\textit{\textbf {Future experiments:}}}\\

Experimentally, important avenues for further investigation are linked to the Fermilab Short Baseline Neutrino (SBN) program~\cite{Bonesini:2022pwy,Machado:2019oxb}.  This consists of three liquid Argon based Time-Projection Chamber detectors fed by the Booster Neutrino Beam (BNB) and includes MicroBooNE. Their overall improved particle-identification capabilities provide greater discriminating power between electron, photon and $e^+e^-$ events. In the near future, data from MicroBooNE on incoherent scattering events (to complement their coherent scattering results discussed above) would help shed light on the nature of possible new physics.  

 Two additional detectors are part of  the SBN program: the ICARUS~\cite{Pia:2025ydd,Roy:2026ntg} detector, currently in operation and with a much larger fiducial mass of 760 t compared to MicroBooNE,  located 600 m from the target, and the SBN Detector~(SBND)~\cite{Tufanli:2016hyo,Dutta:2026dvg} with a 112-t active volume with a shorter baseline of  110 m. Both experiments could  be sensitive to scalar and pseudoscalar mediators in the mass range relevant to the LSND and MB anomalies. They could also shed light on the presence of HNLs that are a feature of many of the models proposed to explain the electron-like excesses discussed here.

\begin{figure}[!t]
\begin{center}
\includegraphics[width=7.5cm,height=6.cm]{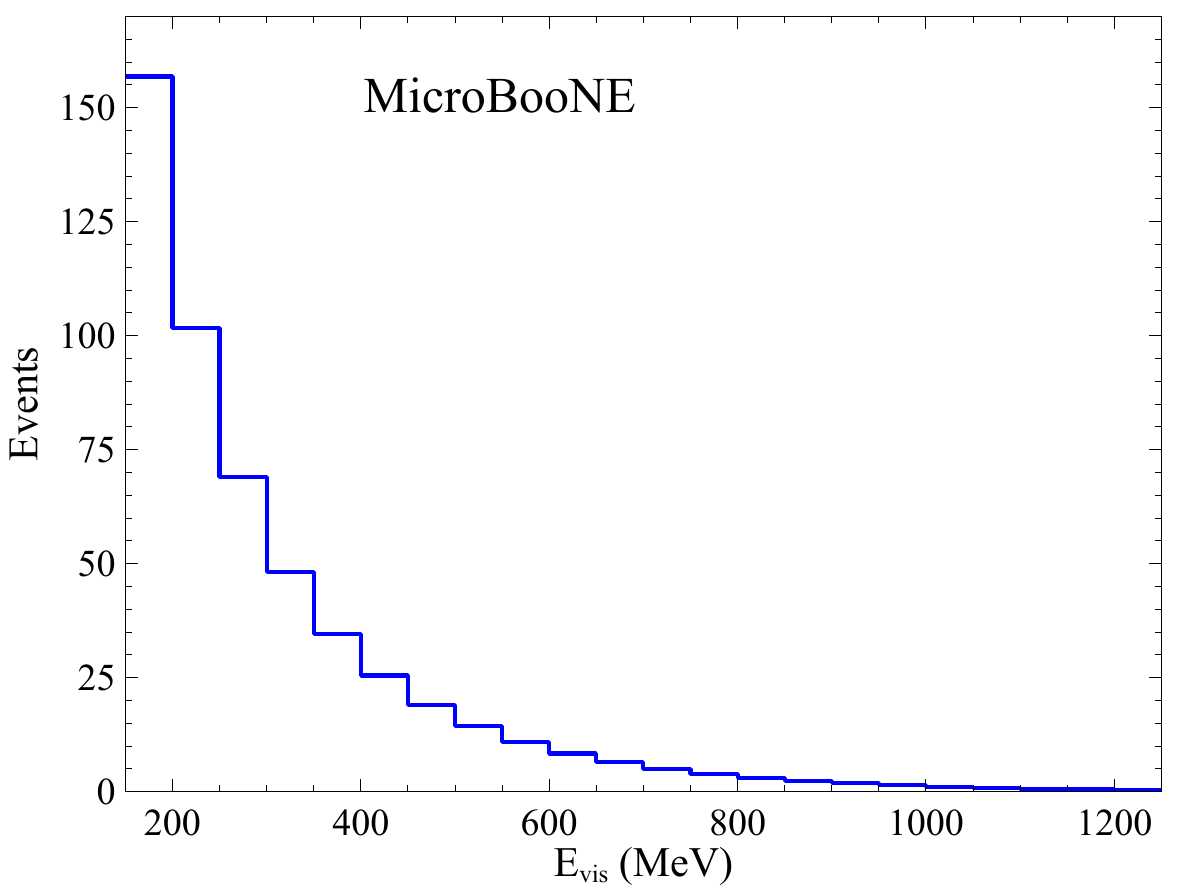}~~~\includegraphics[width=7.5cm,height=6cm]{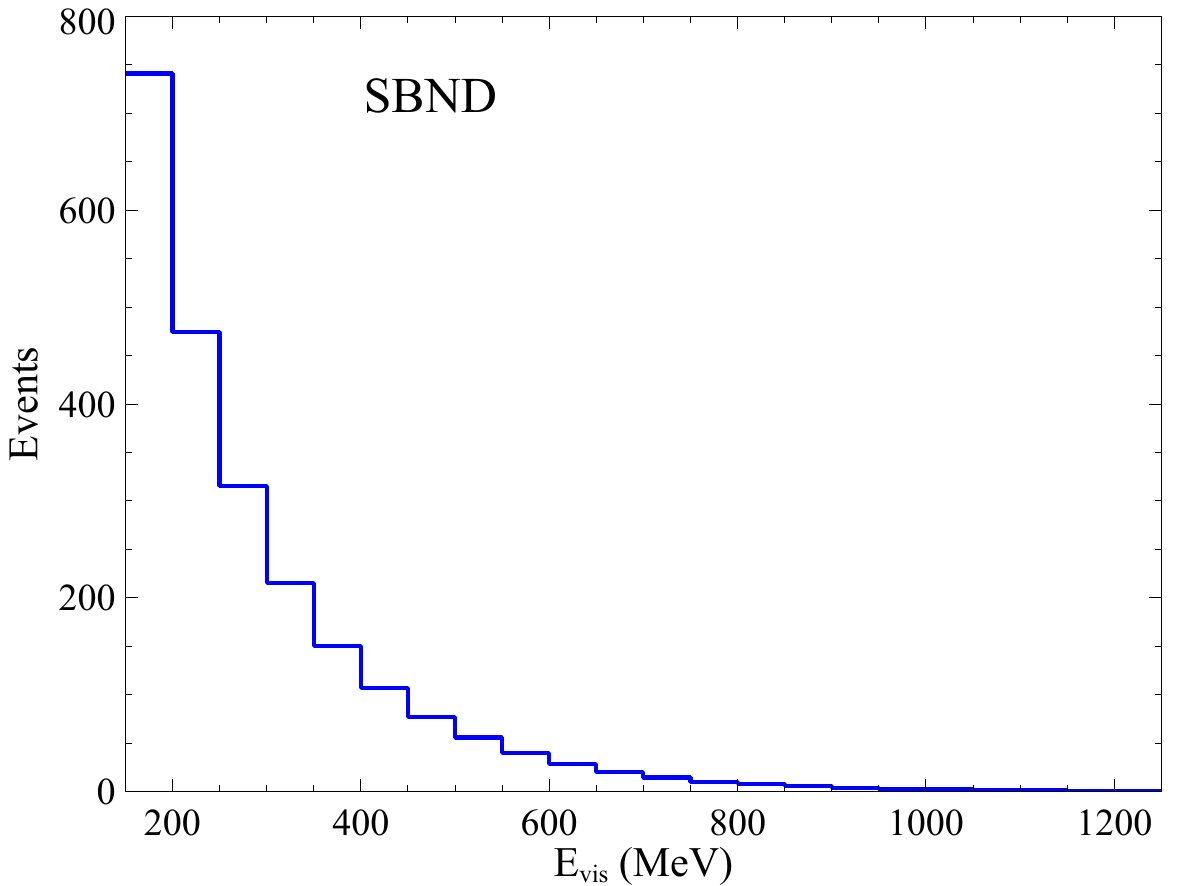}
\caption{Prediction of incoherent events in MicroBooNE and SBND.}
\label{Fig:uB-SBND}
\end{center}
\end{figure}

Using the values in Table~\ref{tab1}, we show the predicted number of incoherent events as a function of the visible energy in figure~\ref{Fig:uB-SBND}. We have assumed a detection efficiency of $12\%$ for an $e^+e^-$ pair in both experiments. The corresponding POT for MicroBooNE (SBND) is $6.34 \times 10^{20}$ ($15.6 \times 10^{20}$). As expected, the number of events at SBND is significantly larger than that at MicroBooNE.

\section{Summary}

If we assume that the anomalies observed in MB and LSND are genuine, by a process of systematic data collection and elimination aligned with the best traditions of empirical physics, these experiments along with MicroBooNE may have brought us to the cusp of new physics discoveries. In this work, we have taken stock of recent results and examined how proposed new physics scenarios stack up against them. These scenarios involve, in various combinations,  new mediators (vectors or scalars), HNLs, and dark portals. They attempt to explain the MB and LSND excesses by final states which include electrons, photons or $e^+e^-$ pairs.  We have argued that the overall data from MicroBooNE favour scalar mediators over vectors both from the point of view of explaining data as well as evading constraints. They also point to new interactions which lead to incoherent scattering being dominant over coherent scattering.

The precise structure of a new physics model involving scalar mediators that explains the MB and LSND excesses may vary in the details of its gauge structure, its full particle content, its dark portal interactions and its UV completion, and must be left to future experiments. However, an important conclusion that seems to emerge from the data is that a light ($m\leq 30$~MeV)\footnote{In our work, we have chosen a benchmark value of 17~MeV for the mass. This choice, besides leading to good fits to the data, is motivated by suggested resolutions of the ATOMKI anomaly~\cite{Krasznahorkay:2015iga,Krasznahorkay:2019lyl,Krasznahorkay:2021joi,Krasznahorkay:2022pxs} and also by a  result from the T2K experiment near detector, ND280; see figure 11 in~\cite{T2K:2020lrr}.} dark neutral spin zero  mediator  with dominantly pseudoscalar couplings, decaying in the detectors to $e^+e^-$ pairs, can account for the observations. In addition, HNLs with masses $\sim {\cal O}(100)$~MeV could be part of the new physics particle spectrum.  We have given an example of a  model with these features (in addition to those in~\cite{Abdallah:2020vgg, Abdallah:2024uby}) which fits the MB and LSND energy and angular distributions and appears to be in consonance with existing constraints.  We have provided a gauge invariant supporting structure by adding a 2HDM  to demonstrate feasibility, and discussed various constraints and issues related to its CP properties and  UV completion. 

  We have indicated that if such scenarios are actually realised in nature, it is possible that neutrinos are Dirac particles. Finally, we have discussed avenues of future   experimental investigation which could provide a test of the ideas presented here and/or lead to a resolution of the anomalies. In the immediate future, such a test would involve a search for the excess in incoherent scattering channels in MicroBooNE and SBND.
\section*{Acknowledgments} 
Raj Gandhi is thankful to William Louis and Mark Ross-Lonergan for useful  exchanges on  MB, LSND and MicroBooNE, and to Anil Thapa for an especially helpful discussion. He also acknowledges helpful conversations with K.S. Babu, P.S. Bhupal Dev, Bhaskar Dutta,  Andre de Gouvea, Matheus Hostert, Sudip Jana, Doojin Kim,  Danny Marfatia and  Takuya Okawa. He is grateful to the to the the Center for Theoretical Underground Physics and Related Areas (CETUP), The Institute for Underground Science at Sanford Underground Research Facility (SURF), and the South Dakota Science and Technology Authority for hospitality and financial support, as well as for providing a stimulating environment   facilitating portions of this research. He also acknowledges support from the Department of Atomic Energy, Government of India, and the  Harish-Chandra Research Institute. The work of WA is partially supported by the Science, Technology and Innovation Funding Authority (STDF) under grant number 50806.
\section{Appendix}
\label{app}
We use the momentum assignment
\[
\nu_\mu(k_1) + n(p_1) \to N(k_2) + n(p_2) + a'(k),
\]
and define the two virtual scalar momenta
\[
q_\nu \equiv k_1 - k_2 \quad(\text{neutrino-side}), \qquad
q_n    \equiv p_1 - p_2 \quad(\text{nucleon-side}),
\]
so momentum conservation at the cubic vertex reads \(q_\nu + q_n = k\).\\

\subsection*{Relevant Feynman rules (all momenta incoming):}
\begin{itemize}
 \item $a' \bar q q$: \quad $-i y^{a'}_q \gamma_5$.
 \item $a' \bar N \nu_\mu$: \quad $-i y^{a'}_{\nu_\mu N} \gamma_5 P_L$ .
 \item $a'^3$: \quad $-i\kappa$.
 \item Pseudoscalar propagator: \quad $\dfrac{i}{q^2 - M_{a'}^2}$.
\end{itemize}
\subsection*{Momentum assignment and masses:}
\[
\begin{aligned}
\text{Incoming } \nu_\mu:\quad&k_1^\mu=(E_{\nu_\mu},\; \vec{k}_1)\simeq(|\vec{k}_1|,\vec{k}_1), \text{that is}\quad k_1^2=m_{\nu_\mu}^2\simeq0.\\[0.3cm]
\text{Initial nucleon}:\quad&p_1^\mu=(E_{p_1},\; \vec p_1)=(m_n,\mathbf0),\qquad p_1^2=E_{p_1}^2=m_n^2.\\[0.3cm]
\text{Outgoing}N:\quad&k_2^\mu=(E_N,\; \vec k_2),\qquad E_N^2-|\vec{k}_2|^2=k_2^2=M_N^2~(\text{on-shell~condition}).\\[0.3cm]
\text{Final nucleon}:\quad&p_2^\mu=(E_{p_2},\; \vec p_2),\qquad E_{p_2}^2-|\vec{p}_2|^2= p_2^2=m_n^2~(\text{on-shell~condition}).\\[0.3cm]
\text{Outgoing}a':\quad&k^\mu=(E_{a'},\; \vec k\,),\qquad E_{a'}^2-|\vec{k}|^2= k^2=M_{a'}^2~(\text{on-shell~condition}).
\end{aligned}
\]

\subsection*{Kinematic relations:}
In the laboratory frame with the incoming muon neutrino along $z$-axis:
\begin{eqnarray}
k_1^\mu&=&(E_{\nu_\mu},0,0,E_{\nu_\mu}),\\
k_2^\mu&=&(E_N,|\vec k_2|\sin\theta_{N},0,|\vec k_2|\cos\theta_N),
\end{eqnarray}
where $\theta_N$ denotes the angle between the directions of the momenta $\vec k_1$ and $\vec k_2$ which lie in the $xz$-plane, therefore,
\begin{equation}
k_1\cdot k_2 = E_{\nu_\mu} E_N - E_{\nu_\mu} |\vec k_2|\cos\theta_N=E_{\nu_\mu} (E_N - \sqrt{E_N^2-M_N^2}\cos\theta_N)\,.
\end{equation}
Also,
\begin{equation}
p_1\cdot p_2 = E_{p_1} E_{p_2}= m_n E_{p_2}\,,~~~p_1\cdot k_1=E_{p_1} E_{\nu_\mu}=m_n E_{\nu_\mu}\,.
\end{equation}
4-momentum conservation:
\begin{equation}
k_1^\mu+p_1^\mu=k_2^\mu+k^\mu+p_2^\mu\,,
\end{equation}
hence,
\begin{equation}
 \vec{p_2}=\vec k_1-\vec k_2 -\vec k~~~ {\rm and}~~~~ E_{p_2}=\sqrt{m_n^2+|\vec{p}_2|^2}~~\to~~E_{p_2}=\sqrt{m_n^2+|\vec k_1-\vec k_2 -\vec k|^2}\,.
\end{equation}
\subsection*{Momentum transfers:}
\begin{eqnarray}
q_\nu^2&=&(k_1-k_2)^2 = m_{\nu_\mu}^2 + M_N^2 -2k_1\cdot k_2 \simeq M_N^2 - 2k_1\cdot k_2,\\
q_n^2&=&(p_1-p_2)^2 = m_n^2 + m_n^2 - 2 p_1\cdot p_2 = 2m_n^2 - 2 p_1\cdot p_2= 2m_n (m_n - E_{p_2}).
\end{eqnarray}
Using these, the tree-level amplitude corresponding to the Feynman diagram in figure~\ref{Fig:FD-MB} is
\begin{equation}\label{eq:amp_iM}
\begin{aligned}
i\mathcal{M}
&= \big[ \bar u_N(k_2)\,(i y^{a'}_{\nu_\mu N} P_L)\, u_\nu(k_1)\big]
\; \frac{i}{\,q_\nu^2 - M_{a'}^2\,} \;
(-i\kappa)\;
\frac{i}{\,q_n^2 - M_{a'}^2\,} \\
&\qquad\qquad \times \; \big[ \bar u_q(p_2)\,(-i y^{a'}_q\gamma_5)\, u_q(p_1)\big] .
\end{aligned}
\end{equation}
Collecting the coupling factors, we get
\begin{equation}\label{eq:amp_M}
\mathcal{M} \;=\;\frac{\,y^{a'}_{\nu_\mu N}\,y^{a'}_q\,\kappa}
{(q_\nu^2-M_{a'}^2)\,(q_n^2-M_{a'}^2)}\;
\big[\bar u_N(k_2) P_L u_\nu(k_1)\big]\;
\big[\bar u_q(p_2)\gamma_5 u_q(p_1)\big].
\end{equation}
To get $\frac{1}{2}\displaystyle\sum_{\rm spins}|\mathcal{M}|^2$, we compute the spin traces:
\begin{eqnarray}\label{Eq:M2}
&&\!\!\!\!\!\!\!\!\!\!\!\!\frac{1}{2}\sum_{\rm spins}|\mathcal{M}|^2=\frac{1}{2}\frac{|y^{a'}_{\nu_\mu N}\,y^{a'}_q\,\kappa|^2}
{\left[(q_\nu^2-M_{a'}^2)\,(q_n^2-M_{a'}^2)\right]^2}\nonumber\\[0.2cm]&&\times\sum_{\rm spins}\big[\bar u_N(k_2) P_L u_\nu(k_1)\bar u_\nu(k_1)P_R u_N(k_2)\big]\big[-\bar u_q(p_2)\gamma_5 u_q(p_1)\bar u_q(p_1)\gamma_5 u_q(p_2)\big]\nonumber\\[0.2cm]
&=& -\frac{1}{2}\frac{|y^{a'}_{\nu_\mu N}\,y^{a'}_q\,\kappa|^2}
{\left[(q_\nu^2-M_{a'}^2)\,(q_n^2-M_{a'}^2)\right]^2}\Tr\big[(\slashed{k}_2+M_N)P_L\,\slashed{k}_1\,P_R\big]\Tr\big[(\slashed{p}_2+m_n)\gamma^5\,[(\slashed{p}_1+m_n)\gamma^5\big]\nonumber\\[0.2cm]
&=& -\frac{1}{2}\frac{|y^{a'}_{\nu_\mu N}\,y^{a'}_q\,\kappa|^2}
{\left[(q_\nu^2-M_{a'}^2)\,(q_n^2-M_{a'}^2)\right]^2}(2k_1\cdot k_2)(4m_n^2-4p_1\cdot p_2)\nonumber\\[0.2cm]
&=& \frac{4|y^{a'}_{\nu_\mu N}\,y^{a'}_q\,\kappa|^2(k_1\cdot k_2)(p_1\cdot p_2-m_n^2)}
{\left[(q_\nu^2-M_{a'}^2)\,(q_n^2-M_{a'}^2)\right]^2}\,,
\end{eqnarray}
here we have used that $m_\nu\simeq0$. The differential cross-section is given by

\begin{equation}
d\sigma=\frac{1}{4E_{\nu_\mu}E_{p_1} v_{\rm rel}}\overline{|\mathcal{M}|^2}\;d\Phi_3(k_1+p_1;k_2,k,p_2),
\end{equation}
where the spin-summed squared amplitude $\overline{|\mathcal{M}|^2}$ is given in eq.~(\ref{Eq:M2}), $$v_{\rm rel}=\frac{\sqrt{(p_1\cdot k_1)^2-m_{\nu_\mu}^2m_n^2}}{E_{p_1}E_{\nu_\mu}}\simeq\frac{\sqrt{(E_{p_1}E_{\nu_\mu})^2}}{E_{p_1}E_{\nu_\mu}}=1,$$  and the 3-body phase space is given by
\begin{equation}
d\Phi_3=(2\pi)^4\delta^4(k_1+p_1-k_2-k-p_2){\frac{d^3\vec k_2}{(2\pi)^3(2E_N)}}{\frac{d^3\vec k}{(2\pi)^3(2E_{a'})}}{\frac{d^3\vec p_2}{(2\pi)^3(2E_{p_2})}}\,.
\end{equation}
Therefore,
\begin{eqnarray}
d\Phi_3&=&\delta^4(k_1+p_1-k_2-k-p_2)\frac{d^3\vec k_2\, d^3\vec k\, d^3\vec p_2}{8 (2\pi)^5 E_N  E_{a'} E_{p_2} } \\
&=&\frac{d^3\vec k_2\, d^3\vec k}{4 (2\pi)^5 E_N  E_{a'} }\, d^4 p_2\;\delta^4(k_1+p_1-k_2-k-p_2)\;\delta(p_2^2-m^2_n),
\end{eqnarray}
where we have used that $d^3\vec p_2=2 E_{p_2} d^4p_2\;\delta(p_2^2-m^2_n)\,H(E_{p_2})$ with the Heaviside step function $H(E_{p_2}>0)=1$ for a physical particle of positive energy, otherwise $d\sigma=0$.

Performing the $d^4p_2$ integral using the 4D delta, we get
\begin{eqnarray}
d\Phi_3&=&\frac{d^3\vec k_2\, d^3\vec k}{4 (2\pi)^5 E_N  E_{a'} }\,\delta((k_1+p_1-k_2-k)^2-m^2_n),
\end{eqnarray}
where the 4D delta sets $p_2^\mu=k_1^\mu+p_1^\mu-k_2^\mu-k^\mu$.

Now, we use:
\begin{equation}
\frac{d^3\vec k_2}{E_N}=|\vec k_2|\, dE_N\, d\Omega_N,~~~~\frac{d^3\vec k}{E_{a'}}=|\vec k|\, dE_{a'}\, d\Omega_{a'},
\end{equation}
thus, 
\begin{equation}
d\Phi_3=\frac{|\vec k_2|\,|\vec k|}{4 (2\pi)^5}\, dE_N\, d\Omega_N\, dE_{a'}\, d\Omega_{a'}\,\delta(F),
\end{equation}
where $F=(k_1+p_1-k_2-k)^2-m^2_n$.

We define the total incoming four-momentum as
\begin{equation}
P^\mu=k_1^\mu+p_1^\mu,
\end{equation}
and work temporarily in the overall center-of-mass (CM) frame, where
\begin{equation}
P^\mu=(\sqrt{s},\vec{0}\,).
\end{equation}
Therefore, in the CM frame,
\begin{eqnarray}
F&=&(P-k_2-k)^2-m^2_n\\
&=&
s+M_N^2+M_{a'}^2-m_n^2
+2E_NE_{a'}
-2\sqrt{s}(E_N+E_{a'})
-2|\vec{k}_2|\,|\vec{k}|
\cos\theta_{Na'},
\end{eqnarray}
where $\theta_{Na'}$ is the angle between the directions of the momenta $\vec k_2$ and $\vec k$ which lie in the $xz$-plane. Thus,
\begin{eqnarray}
d\Phi_3&=&\frac{|\vec k_2|\,|\vec k|}{4 (2\pi)^5}\, dE_N\, d\Omega_{a'}\, dE_{a'}\, d\phi_{Na'}d(\cos{\theta_{Na'}})\,\frac{\delta(\cos{\theta_{Na'}}-\cos{\theta^*_{Na'}})}{2|\vec{k}_2|\,|\vec{k}|}~~~~~~~~\\
&=&\frac{1}{8 (2\pi)^5}\, dE_N\, d(\cos\theta_{a'})\,d\phi_{a'}\, dE_{a'}\, d\phi_{Na'},
\end{eqnarray}
where
\begin{equation}
\cos{\theta^*_{Na'}}=\frac{s+M_N^2+M_{a'}^2-m_n^2
+2E_NE_{a'}
-2\sqrt{s}(E_N+E_{a'})}{2|\vec{k}_2|\,|\vec{k}|}\,.  
\end{equation}
Therefore,
\begin{equation}
d\Phi_3=\frac{1}{8 (2\pi)^5}\,d\phi_{a'}\, d\phi_{Na'}\,\frac{|\vec{k}|_{\rm lab}}{|\vec{k}|}\, d(\cos\theta^{\rm lab}_{a'})\, dE^{\rm lab}_{a'}\, dE_N,
\end{equation}
where
\begin{equation}
E_{a'}
=
\gamma
\left[
E_{a'}^{\rm lab}
-\beta
|\vec{k}|_{\rm lab}
\cos\theta_{a'}^{\rm lab}
\right],~~~~~~
\cos\theta_{a'}
=
\frac{\gamma}{|\vec{k}|}\left[
|\vec{k}|_{\rm lab}\cos\theta_{a'}^{\rm lab}
-\beta E_{a'}^{\rm lab}
\right],
\end{equation}
with
\begin{equation}
|\vec{k}|_{\rm lab}=\sqrt{(E^{\rm lab}_{a'})^2-M^2_{a'}},~~~~\gamma^2=\frac{1}{1-\beta^2}=\frac{(E_{\nu_\mu}+m_n)^2}{s}.
\end{equation}    
Thus, the differential cross-section with respect to $E^{\rm lab}_{a'}$ can be written as
\begin{equation}
\frac{d\sigma}{dE_{a'}^{\rm lab}}
\!=\!
\frac{1}
{32(2\pi)^4m_nE_{\nu_\mu}}
\!\int_0^{2\pi}\!\!d\phi_{Na'}\!\!\int_{-1}^{1} \!\!d(\cos\theta^{\rm lab}_{a'})\!\!
\int_{E_N^{\min}(E_{a'})}^{E_N^{\max}(E_{a'})}
dE_N
\frac{|\vec{k}|_{\rm lab}}{|\vec{k}|}\left.\overline{|\mathcal{M}|^2}\right|_{\cos\theta_{Na'}=\cos\theta_{Na'}^*},
\end{equation}
where
\begin{equation*}
E_N^{\min,\max}(E_{a'})
=
\frac{
(\sqrt{s}-E_{a'})(M_{Nn}^2+M_N^2-m_n^2)
\mp
\sqrt{E_{a'}^{2}-M_{a'}^2}\,
\lambda^{1/2}(M_{Nn}^2,M_N^2,m_n^2)
}
{2M_{Nn}^2},
\end{equation*}
\begin{equation*}
M_{Nn}^2
=
s+M_{a'}^2-2\sqrt{s}\,E_{a'},~~~~
\lambda(x,y,z)
=
x^2+y^2+z^2-2xy-2xz-2yz,
\end{equation*}
with
\begin{equation*}
M_{a'}\leq E_{a'}\leq
\frac{
s+M_{a'}^2-(M_N+m_n)^2
}
{2\sqrt{s}},~~~~~~~~s=m_n^2+2m_n E_{\nu_\mu}.
\end{equation*}

\bibliographystyle{apsrev4-1}
\bibliography{Refs}
\end{document}